%% file: root.tex
\documentclass[letterpaper,journal, 10 pt, twoside]{IEEEtran} 
\input{preamble.tex} 

\usepackage{balance}

	\def\BibTeX{{\rm B\kern-.05em{\sc i\kern-.025em b}\kern-.08em
				T\kern-.1667em\lower.7ex\hbox{E}\kern-.125emX}}
\usepackage{lettrine} 

\title{Revisiting the Optimal PMU Placement Problem \\ in Multi-Machine Power Networks}

\author{Mohamad H. Kazma, \textit{Graduate Student Member, IEEE} and Ahmad F. Tah$\text{a}^{\diamond}$, \textit{Member, IEEE}
	\thanks{
		$^\diamond$Corresponding author.  This work is supported by National Science Foundation under Grants 2152450 and 2151571. The authors are with the Civil $\&$ Environmental Engineering and Electrical $\&$ Computer Engineering Departments, Vanderbilt University, 2201 West End Ave, Nashville, Tennessee 37235. Emails: mohamad.h.kazma@vanderbilt.edu, ahmad.taha@vanderbilt.edu.}
}

\begin{document}
\setlength{\abovedisplayskip}{3.1pt}
\setlength{\belowdisplayskip}{3.1pt}
\setlength{\abovedisplayshortskip}{3.1pt}
\setlength{\belowdisplayshortskip}{3.1pt}
		
\newdimen\origiwspc%
\newdimen\origiwstr%
\origiwspc=\fontdimen2\font
\origiwstr=\fontdimen3\font


\maketitle
\thispagestyle{headings} 
\pagestyle{headings}

\begin{abstract}
		To provide real-time visibility of physics-based states, phasor measurement units (PMUs) are deployed throughout power networks. PMU data enable real-time grid monitoring and control---and are essential in transitioning to smarter grids. Various considerations are taken into account when determining the geographic, optimal PMU placements (OPP). This paper focuses on the control-theoretic, observability aspect of OPP. A myriad of studies have investigated observability-based formulations to determine the OPP within a transmission network. However, they have mostly adopted a simplified representation of system dynamics, ignored basic algebraic equations that model power flows, disregarded including renewables such as solar and wind, and did not model their uncertainty. Consequently, this paper revisits the observability-based OPP problem by addressing the literature's limitations. A nonlinear differential algebraic representation (NDAE) of the power system is considered. The system is discretized using various discretization approaches while explicitly accounting for uncertainty. A moving horizon estimation approach is explored to reconstruct the joint differential and algebraic initial states of the system, as a gateway to the OPP problem which is then formulated as a computationally tractable integer program (IP). Comprehensive numerical simulations on standard power networks are conducted to validate the different aspects of this approach and test its robustness to various dynamical conditions.
\end{abstract}
\begin{IEEEkeywords}
		Power system modeling, phasor measurements, nonlinear differential algebraic models, nonlinear observability, moving horizon estimation, optimal PMU placement
\end{IEEEkeywords}
\section{Introduction and Paper Contributions}\label{sec:Introduction}
\subsection{Introduction and Literature Review}\label{subsec:Intro-lit}
\noindent \lettrine[lines=2]{W}{ithin} recent years, there has been a trend of shifting from fuel-based energy generation to fuel-free, renewable resources. This transition has imposed a challenge on maintaining stable grid operations~\cite{Zhao2012}.
To enable stable and secure power systems operations, grid monitoring is typically performed by the supervisory control and data acquisition (SCADA) system~\cite{Chen2022}. However, SCADA measurements cannot fully capture the dynamics of the power system under transient conditions due to slow system update rates~\cite{Ahmadi2011, Yuill2011, Qi2015}. As such, more elaborate real-time system monitoring is required to perform state estimation and control. Dynamic state estimation (DSE) from real-time measurements allows for reconstructing system states, thereby enabling stability predictions and control. In order to achieve an observable system for DSE, PMUs are installed within the grid, thereby providing real-time high-resolution measurements~\cite{Sun2016}. Albeit a system can have PMUs allocated to each bus and achieve complete observability, this allocation is not economical~\cite{Nugroho2021b}. Thus, it is necessary to solve for optimal PMU placements as a constraint optimization problem whilst achieving observability~\cite{Yuill2011}.

Previous studies have focused on the OPP while typically considering steady-state estimation (SSE).
The study in~\cite{Li2013} proposes an information gain approach that is able to model state uncertainties. The proposed NP-hard problem is solved using a greedy algorithm that achieves an approximate solution. On a similar note,~\cite{Kumar2018} utilizes a reliability based observability criterion to rank PMU locations. In~\cite{Shi2020}, PMU placement is based on the minimization of mean-squared error between measurement output and grid states. In~\cite{RezaeianKoochi2020}, a binary linear optimization program is proposed; it solves the OPP while maintaining full observability and assuring backup protection. Interested readers are referred to~\cite{Manousakis2012, Nazari-Heris2015} for a comprehensive survey of the different approaches that deal with OPP under topological observability for power systems.

We note that the OPP frameworks posed in the aforementioned brief survey are based on topological observability that does not necessarily imply numerical observability~\cite{Sodhi2010,Ghosh2022}. Recently, a two-stage OPP approach has been developed to ensure the numerical observability of the optimal PMU placement obtained from a topological approach; it is based on a rank condition that is qualitative~\cite{Ghosh2022}. 
Such methods that are based on steady-state operating conditions can be unsuitable for power systems that exhibit considerable dynamic changes~\cite{AliAbooshahab2021}. That being said, SSE based sensor placement under a topological observability framework tends to neglect important dynamic parameters, such as system inertia and fluctuating/uncertain load injections~\cite{Qi2015,Korres2003,Sodhi2010,Ghosh2022}.

With that in mind, we focus on formulating the OPP from a numerical observability-theoretic perspective. Meaning that the placement problem is posed so that the system is observable and therefore, control and stability predictions under DSE are achievable~\cite{Balas1999}. A primary step in developing an observability-based OPP framework is to quantify the observability of the dynamic system. This quantification for nonlinear networks can be approached under several formulations. One approach is substituting the nonlinearities with first order linear approximations~\cite{Rouhani2017, Vinod2022a}. This approach yields inaccurate observability-based analyses under uncertainties that change the operating conditions for which the linearized dynamics are valid. Other prevalent approaches include the use of Lie derivatives and differential embedding to construct the observability matrix~\cite{Letellier2005,Whalen2015}. However, such formulations do not guarantee optimal sensor selection for network observability, since they yield measures of observability that are qualitative in nature~\cite{Krener2009,Haber2018}. Furthermore, quantifying observability in nonlinear networks can be based on the empirical observability Gramian~\cite{lall1999empirical,Hahn2002,Krener2009}. The positive semidefinite matrix structure of the empirical observability Gramian, which relates energy notions of controllability and observability quantitative metrics---the trace, determinant, and rank---allows for sensor selection~\cite{Summers2016}. 

A state-of-the-art review that details the different topological and numerical observability-based OPP problems for transmission networks is presented in~\cite{Netto2022}. Howbeit, current literature that has addressed the combinatorial observability-based OPP problem in nonlinear power networks establish that the sensor selection problem $(i)$ is not well understood and is solved via heuristic methods that become infeasible for large networks~\cite{Summers2016} and $(ii)$ is still considered an open problem for a nonlinear representation of the power system~\cite{Haber2018}. This is due to the complexity of the nonlinearities that become evident in the observability analysis of the system.

Reference~\cite{Zhang2010} presents one of the initial studies related to DSE and observability for posing an OPP problem while considering a simplified power system model ($2^{nd}$-order generator model). The resulting mixed-integer OPP problem is evaluated by utilizing stochastic estimates of the steady-state covariance matrix derived from the discrete algebraic Riccati equation. In~\cite{Serpas2013}, the optimal sensor placement for a nonlinear network is posed as a maximization of the empirical observability Gramian's determinant; however, for systems that are marginally observable, this metric results in numerical problems. Similarly for a nonlinear power system,~\cite{Qi2015} poses the OPP problem based on empirical observability Gramian metrics. Although good observability under the optimal placements is obtained, the OPP framework: $(i)$ is posed under typical power flow conditions and then the robustness of the optimal solution is examined; $(ii)$ is limited to estimation of dynamic states while neglecting algebraic state estimation; and $(iii)$ is posed as a nonconvex mixed integer program that can be computationally exhaustive for larger systems. To tackle the computational complexity of the OPP, the authors in~\cite{Summers2016} approach solving the OPP problem by leveraging the observability Gramian metrics' submodularity properties. However, the approach yields suboptimal placements since it is solved using a greedy approach. In~\cite{Rouhani2017}, an accurate observability estimation for a power system was computed using Lie derivatives, however, such method is complex for highly nonlinear systems. Solving the OPP problem under such framework would be infeasible even for small networks.

Furthermore, recent literature related to power systems observability~\cite{Qi2016,Yang2016, Saleh2017,Romao2018,Zheng2021,Dang2022} extend the work developed by~\cite{Qi2015,Summers2016,Rouhani2017}. However, the power system models considered are limited to an ordinary differential equation (ODE) representation. Typically, the differential equations are considered in the system representation of the model, whereas the algebraic equations are often neglected due to the computational burden and overall stability implications~\cite{Liu2020}. A complete representation of a power system includes both differential and algebraic equations. In~\cite{Qi2015,Qi2016,Dang2022} both a simplified $2^{nd}$-order generator model and a more detailed $4^{th}$-order generator model are considered, while in~\cite{Rouhani2017,Zheng2021} a discretized $4^{th}$-order generator model with an IEEE-Type 1 and IEEE-DC1A exciter, respectively, are considered. We note here that the algebraic equations are overlooked in the dynamic modeling of the power system in aforementioned OPP literature. Representing a complete differential and algebraic power system model has gained recent interest within the power systems community, refer to~\cite{Milano2016,Grob2016,Liu2020,Nadeem2021,Nugroho2022,Nugroho2023a}. The advantages of simulating the dynamics of the system under a complete NDAE formulation are: $(i)$ linking of network dynamics with power flow equations resulting in an accurate dynamics representation~\cite{Nugroho2022}, $(ii)$ modeling load and renewable uncertainties in DSE routines as a result of incorporating renewables in the system~\cite{Grob2016}, and $(iii)$ expanding the set of potential measurement buses to include non-generator buses.

\subsection{Paper Contributions and Organization}
Motivated by the aforementioned limitations related to nonlinear observability and modeling of nonlinear power systems, the objective of this work is to develop an OPP framework for an NDAE representation of a power system while $(i)$ achieving full observability under uncertainty from load and renewable injections, $(ii)$ jointly estimating both dynamic and algebraic states of the power systems, and $(iii)$ posing the OPP under a computationally efficient formulation. To the best of our knowledge, observability-based OPP for power systems represented as an NDAE has not yet been investigated. Accordingly, we approach formulating the OPP problem on the basis of leveraging the modularity of the observability matrix. The main contributions of this work are as follows:
\begin{itemize}[leftmargin=*]
	\item We introduce and validate a structure preserving transformation that retains the complete NDAE representation while achieving a nonlinear ODE formulation. By using this model---denoted as $\mu$-NDAE---we show that the proposed system's observability can be computed.
	We consider three different implicit discrete-time modeling methods: backward differential formula (BDF), backward Euler (BE) and trapezoidal implicit (TI).
	\item As a stepping stone for the OPP problem, we reconstruct the joint dynamic and algebraic initial states by adopting a moving horizon estimation (MHE) framework. The state estimation is posed as a nonlinear least-squares problem which we solve numerically using the Gauss-Newton algorithm.
	\item We leverage the modularity property of the observability matrix to pose the OPP as a convex integer program (IP). Based on the modularity of the observability matrix, \textit{a priori} observability information from each PMU placement is extracted prior to solving the OPP. Such approach extenuates the computational complexity of an optimization instance resulting  in a computationally tractable approach for PMU placement.
	\item The validity and effectiveness of this approach are studied on standard power networks. We show the validity of the $\mu$-NDAE model for several discretization methods and we prove that the OPP for a specific number of PMUs are subsets of that for a larger number of PMUs, thus indicating modularity.
\end{itemize}

A preliminary and partial version of this paper appeared in~\cite{Kazma2023} without proofs; it included a small case study on the viability of the proposed OPP approach. In this paper, we include several theoretical and numerical developments by:
$(i)$ extending the proposed approach under three implicit discretization methods,
$(ii)$ providing detailed proofs and explicit Jacobian formulations for building the observability measures for each method, 
and $(iii)$ extending the numerical studies to include OPP for larger power networks.

The remainder of this paper is organized as follows. In Section \ref{sec:nonlinearmodel}, we introduce the NDAE power system and its state-space formulation. In Section \ref{sec:discretenonlinearmodel}, we present the different implicit discretizations of the NDAE system and the $\mu$-NDAE system. In Section~\ref{sec:Initial-State-Estimation}, initial state estimation based on a MHE approach is developed. In Section \ref{sec:OSP-Power-system}, the OPP problem is formulated. The proposed OPP is studied for several standard power networks in Section \ref{sec:case_study}. Finally, Section \ref{sec:summary} concludes the paper.\\
\textit{Notation:}~Let $\mathbb{N}$, $\mathbb{R}$, $\mathbb{R}^n$, and $\mathbb{R}^{p\times q}$ denote the set of natural numbers, real numbers, and real-valued row vectors with size of $n$, and $p$-by-$q$ real matrices respectively. The symbol $\otimes$ denotes the Kronecker product. The cardinality of a set $\mathcal{N}$ is denoted by $|\mathcal{N}|$. The operators $\mathrm{det}(\cdot)$ and $\mathrm{trace}(\cdot)$ return the determinant and trace of a matrix. The operator $\mathrm{blkdiag}(\cdot)$ constructs a block diagonal matrix. The operator $\{\m{x}_{i}\}_{i=0}^{\mr{N}} \in \mathbb{R}^{\mr{N}n}$ constructs a column vector that concatenates vectors $\m{x}_i \in \mathbb{R}^{n}$ for all $i \in \{0, \dots, \mr{N}\}$.
\section{Nonlinear Power System DAE Model}\label{sec:nonlinearmodel}
A power system $(\mathcal{N} ,\mathcal{E})$ can be represented graphically, where $\mathcal{E} \subseteq \mathcal{N} \times \mathcal{N}$ is the set of transmission lines, $\mathcal{N} = \mathcal{G} \cup \mathcal{L} \cup \mathcal{R} $ is the set of all buses in the network. Readers are referred to Tab.~\ref{tab:notation} in Appendix~\ref{apdx:nom} for the definitions and notations used in modeling the power system dynamics presented in~\eqref{eq:differential_dynamics}--\eqref{eq:power_balance}.

In this work, a NDAE formulation of a power system is studied. We consider the standard two-axis $4^{th}$-order transient model of a synchronous generator~\cite{Sauer2017}. This model excludes the exciter dynamics, meaning that each of the generators has 4 states and 2 control inputs. We note here that the power system model is controlled by the governor response~\eqref{eq:differential_dynamics_4} by changing the active power produced by a generator. Readers can refer to~\cite[Ch. 7]{Sauer2017} for additional description of the power network utilized within this work. The dynamics of a synchronous generator $i \in \mathcal{G}$ can be written as 
\begin{subequations}\label{eq:differential_dynamics}
	\begin{align}
		&\hspace{-0.4cm}\dot{\delta_{i}} =  \omega_{i} - \omega_{0},\label{eq:differential_dynamics_1}\\
		&\hspace{-0.4cm}M_i\dot{\omega}_{i} = T_{\mr{M}i} - P_{\mr{G}i} - D_{i}(\omega_{i}-\omega_{0}),\label{eq:differential_dynamics_2}\\
		&\hspace{-0.4cm}T^{'}_{\mr{d0}i} \dot{E^{'}_{i}} = -\frac{x_{\mr{d}i}}{x^{'}_{\mr{d}i}}E^{'}_{i} + \frac{x_{\mr{d}i}- {x_{\mr{d}i}^{'}}}{x^{'}_{\mr{d}i}}v_{i}\cos(\delta_{i}- \theta_{i}) + E_{\mr{fd}i},\label{eq:differential_dynamics_3}\\
		&\hspace{-0.4cm}T_{\mr{CH}i}\dot{T}_{\mr{M}i} = T_{\mr{M}i} - \frac{1}{R_{\mr{D}i}}(\omega_{i}-\omega_{0}) + T_{\mr{r}i},\label{eq:differential_dynamics_4}
	\end{align}
\end{subequations}
where the time varying components in \eqref{eq:differential_dynamics} are: the generator internal states $\delta_{i}$, $\omega_{i}$, ${E}^{'}_{i}$, ${T}_{\mr{M}i}$ and the generator inputs $E_{\mr{fd}i}$, $T_{\mr{r}i} $.

The constants in \eqref{eq:differential_dynamics} are: $M_{i}$, $D_{i}$, $x_{\mr{d}i}$, $x_{\mr{q}i}$, $x^{'}_{\mr{d}i}$, $T^{'}_{\mr{d0}i}$, $T_{\mr{CH}i}$, $R_{\mr{D}i}$, and $\omega_0$. Note that this generator model does not include the dynamical equation for the exciter; modeling its dynamics results in a $9^{th}$-order model that includes a set point for field voltage ${E}_{\mr{fd}i}$ using an automatic voltage regulator.

The algebraic constraints of the power system represent the relation between the internal states of a synchronous generator and its generated power output $P_{\mr{G}i}$ and $Q_{\mr{G}i}$ i.e, real and reactive power. The algebraic constraints of the nonlinear descriptor system can be written as~\eqref{eq:algebriac_constraints} with $i \in \mathcal{G}$.

\begin{subequations}\label{eq:algebriac_constraints}
	\begin{align}
		\hspace{-0.2cm} P_{\mr{G}i} =& \tfrac{1}{x^{'}_{\mr{d}i}}E^{'}_{i}v_{i}\sin(\delta_{i}-\theta_{i})  - \tfrac{x_{\mr{q}i}-x^{'}_{\mr{d}i}}{2x^{'}_{\mr{d}i}x_{\mr{q}i}}v^{2}_{i}\sin(2(\delta_{i}-\theta_{i})),\label{eq:algebriac_constraints_1}\\
		\begin{split}
		\hspace{-0.2cm} Q_{\mr{G}i} =& \tfrac{1}{x^{'}_{\mr{d}i}}E^{'}_{i}v_{i}\cos(\delta_{i}-\theta_{i}) - \tfrac{x_{\mr{q}i}-x^{'}_{\mr{d}i}}{2x^{'}_{\mr{d}i}x_{\mr{q}i}}v^{2}_{i}\\
		&-\tfrac{x_{\mr{q}i}-x^{'}_{\mr{d}i}}{2x^{'}_{\mr{d}i}x_{\mr{q}i}}v^{2}_{i}\cos(2(\delta_{i}-\theta_{i})).
		\end{split}\label{eq:algebriac_constraints_2}
	\end{align}	
\end{subequations}

The power balance between the set of generator, load buses and renewable energy sources, with $i \in \mathcal{G} \cup \mathcal{L} \cup \mathcal{R}$, can be written as~\eqref{eq:power_balance} such that, $N := |\mathcal{N}|$ is the number of buses within the transmission network while, $G := |\mathcal{G}|$, $L := |\mathcal{L}|$, and  $R := |\mathcal{R}|$ are the number of generator, load and renewable buses. Note that we are modeling renewables to be injective into the network, i.e., that renewables are modeled as negative loads.

\begin{subequations}\label{eq:power_balance}
	\begin{align}
		&\hspace{-0.2cm}P_{\mr{G}i} + P_{\mr{L}i} +  P_{\mr{R}i} = \sum_{j=1}^{N} v_{i}v_{j}(G_{ij}\cos\theta_{ij}+B_{ij}\sin\theta_{ij}),\\
		&\hspace{-0.2cm}Q_{\mr{G}i} +Q_{\mr{L}i} +Q_{\mr{R}i} =  \sum_{j=1}^{N} v_{i}v_{j}(G_{ij}\cos\theta_{ij}-B_{ij}\sin\theta_{ij}),
	\end{align} 
\end{subequations}
where $\theta_{ij} = \theta_{i} - \theta_{j}$, i.e, the bus angle. Note that the absence of any generator, load, and renewable at node $i$ can be indicated by setting the respective active and reactive power in~\eqref{eq:power_balance} to zero.

Having presented~\eqref{eq:differential_dynamics}--\eqref{eq:power_balance}, which depict the physics based components of the electromechanical transients---representing both the generator dynamics and algebraic constraints---the state-space formulation of the power system can be written as
\begin{subequations}\label{eq:semi_NDAE_rep}
	\begin{align}
		\textit{generator dynamics}:	\;\; \dot{\m x}_{d} &=  \m{f}(\m x_d,\m x_a, \m u), \label{X_d} \\
		\textit{algebraic constraints}:	\;\; \m 0 & = \m{g}(\m x_d, \m x_a), \label{X_a}
	\end{align} 
\end{subequations}
where the dynamic states of the synchronous machine can be defined as $\m{x}_{d} := \m{x}_{d}(t) =  [\m \delta^{\top} \; \m \omega^{\top} \; {\mE^{'}}^{\top} \; {\m{T}_\mr{M}}^{\top} ]^{\top} \in \mathbb{R}^{4G}$, where $\m{\delta} := \{\delta_{i}\}_{i=0}^{G}$, $\m{\omega} := \{\omega_{i}\}_{i=0}^{G}$, $\m{E^{'}} := \{E^{'}_{i}\}_{i=0}^{G}$, and $\m{T}_\mr{M} := \{{T}_{\mr{M}i} \}_{i=0}^{G}$. The algebraic states can be defined as $\m{x}_{a} := \m{x}_{a}(t) = [\m{P}_\mr{G}^{\top}\; \m{Q}_\mr{G}^{\top} \; \m v^{\top} \; \m\theta^{\top}]^{\top} \in \mathbb{R}^{2G+2N}$, where $\m{P}_\mr{G} := \{{P}_{\mr{G}i} \}_{i=0}^{G}$, $\m{Q}_\mr{G} := \{{Q}_{\mr{G}i} \}_{i=0}^{G}$,
$\m{v} := \{{v}_{i} \}_{i=0}^{N}$, and $\m{\theta} := \{{\theta}_{i} \}_{i=0}^{N}$.
The input of the system can be defined as $\m{u}:=\m{u}(t) = [\m{E}_\mr{fd}^{\top} \; \m{T}_\mr{r}^{\top}]^{\top} \in \mathbb{R}^{2G}$, where $\m{E}_\mr{fd}:= \{{E}_{\mr{fd}i} \}_{i=0}^{G}$ and $\m{T}_\mr{r}:= \{{T}_{\mr{r}i} \}_{i=0}^{G}$. Matrix functions $\m{f}(\cdot)$ and $\m{g}(\cdot)$ are nonlinear mapping functions such that, $\m{f}(\cdot): \mathbb{R}^{4G}\times\mathbb{R}^{2G}\times\mathbb{R}^{2G}\rightarrow\mathbb{R}^{4G}$ and $\m{g}(\cdot): \mathbb{R}^{4G}\times\mathbb{R}^{2G}\times\mathbb{R}^{2N}\rightarrow\mathbb{R}^{2G+2N}$.

\section{Discrete-time Modeling of Power Systems}\label{sec:discretenonlinearmodel}
In this section, we introduce the NDAE formulation referred to as the $\mu$-NDAE system. The impact of the choice of discretization method on the solution to the OPP is unclear. To that end, we investigate several implicit discrete-time modeling techniques and embed them within the OPP formulation presented in Section~\ref{sec:OSP-Power-system}.

DAE solvability has been thoroughly presented and investigated in literature. MATLAB is capable of solving DAEs with the DAE solvers---$\mr{ode15i}$ and $\mr{ode15s}$~\cite{Shampine1999}. However, given the discrete-time modeling approach that the observability-based OPP is herein based on, we refer to the use of numerical methods for simulating the NDAE power system. DAEs are considered unequivocally stiff and in particular, nonlinear power system models exhibit stiff dynamics~\cite{Astic1994b}. Implicit discretization methods when used to simulate stiff dynamics offer a stable and computational efficient solution as compared with explicit discretization techniques. Implicit methods previously used in the context of discrete-time modeling of power systems include: backward differential formulas (BDF) known as Gear's method~\cite{Gear1971}, implicit Runge-Kutta (IRK) method~\cite{Ascher1991,Haber2018}
and trapezoidal implicit (TI) method~\cite{alma, Potra2006}. The IRK method is the most numerically involved. BDF and TI methods have been shown to be efficient methods for simulating power systems for transient stability analysis~\cite{Milano2016}.
\subsection{Discrete-time Representation of NDAEs}\label{subsec:disc_NDAE}
We investigate the use of three implicit time-modeling methods (backward Euler (BE), TI, and BDF) for solving the dynamics of system \eqref{eq:semi_NDAE_rep}.
Such methods account for the stiffness and complexity of the NDAEs. Solving NDAEs using implicit numerical techniques requires finding a solution to a set of implicit nonlinear equations, which we implement using the Newton-Raphson (NR) method. In this section, we will showcase the discrete-time modeling approach for Gear’s method.

Gear's $k_g$-step discretization method is generally stable for BDF discretization index $k_g$ in the range of $2\leq k_g \leq 5$. For $k_g = 1$, Gear's method reduces to BE. Thus, BDF discretization method is a generalization of BE discretization. Accordingly, the discrete-time representation of~\eqref{eq:semi_NDAE_rep} under Gear's method can be written as~\eqref{eq:BDF-discrete_NDEA_semi} for time-step $k$ with step-size $h$, such that $\m{x}_{k}:=\m{x}_{kh}$. We define vectors $\m{z}_{k} :=[\m{x}_{d,k},\; \m{x}_{a,k},\; \m{u}_{k}]^{\top}$ and $\m{x}_{k} :=[\m{x}_{d,k},\; \m{x}_{a,k}]^{\top}$ for time-step $k$, and BDF discretization constant as $\tilde{h} = \beta h$.
\begin{equation}\label{eq:BDF-discrete_NDEA_semi}
	\centering 
	\m{x}_{d,k} - \sum_{s=1}^{k_g}\alpha_s \m {x}_{d,k-s}=  \tilde{h}  \m{f}(\m{z}_{k}), \quad \quad 
	\m{0}  = \m{g}(\m{x}_{k}),
\end{equation}
where the term $\sum_{s=1}^{k_g} \alpha_s \m{x}_{d,k-s}$ represents $k_g$ previous time-steps, and discretization constants $\beta$ and $\alpha_s$, that depend on order of index $k_g$, are calculated as 
\begin{equation}\label{eq:beta_alpha}
	\beta = \Big(\sum_{s=1}^{k_g} \frac{1}{s}\Big)^{-1}  \;,\;\;  \alpha_s = (-1)^{(s-1)} \beta \sum_{j=s}^{k_g} \frac{1}{j}  \begin{pmatrix}
		j \\ s
	\end{pmatrix}.
\end{equation}
\subsection{Structure-preserving $\mu$-NDAE Representation}\label{subsec:ODE}
Before introducing the methodology that is used to solve the NDAE system, i.e., descriptor system, we present a mathematical structural transformation to the NDAE. This involves transforming the system in~\eqref{eq:semi_NDAE_rep} from a NDAE into a nonlinear ODE representation. A descriptor system's index plays an important role in the complexity of the numerical simulation, whereby the higher an index, the more difficult it is to run the system~\cite{Grob2016}. The index of the NDAE system is related to its algebraic equations and refers to the overall equivalency a NDAE has to an ODE~\cite{Gross2014}. The index-n of a NDAE system can be defined as
\begin{mydef}\label{def:inde(1)}
	The descriptor system \eqref{eq:differential_dynamics}-\eqref{eq:power_balance} is said to be of index-1 if the DAEs can be converted into a system of ODEs by differentiating the system with respect to independent time variable $t$ only once. That being said, the index-n of the descriptor system is the number of times needed to differentiate the system of DAEs to obtain a system of ODEs.
\end{mydef}
For the descriptor system \eqref{eq:differential_dynamics}-\eqref{eq:power_balance}, it can be shown that the system is of index-1~\cite{Milano2016, Nugroho2022,Nugroho2023a}. This means that in order to resolve the algebraic constraints into an ODE form, only one differentiation is required. The implicit function theorem can be used to transform system~\eqref{eq:semi_NDAE_rep} from a DAE to an ODE structure~\cite{Krantz2013}.  Applying the aforementioned theorem and differentiating \eqref{X_a} with respect to time, we obtain a NDAE model that is structurally equivalent to an ODE model that is written as
\begin{subequations}\label{eq:NDAE_to_ODE}
		\begin{align}
			\dot{\m{x}}_d &= \m{f}(\m{x}_d,\m{x}_a,\m{u}),\label{eq:NDAE_to_ODE_1} \\
			\dot{\m{x}}_a  &= \tilde{\m{g}}(\m{x}_d,\m{x}_a,\m{u})  = - (\mG_{\m{x}_a})^{-1}\mG_{\m{x}_d}\m f(\m{x}_d,\m{x}_a,\m{u}), \label{eq:NDAE_to_ODE_2} 
		\end{align}
\end{subequations}
where matrix $\mG_{\m{x}_a} =\frac{\partial \m g(\m{x}_d,\m{x}_a) }{\partial \m{x}_a}$ and matrix $\mG_{\m{x}_d} =\frac{\partial \m g(\m{x}_d,\m{x}_a) }{\partial \m{x}_d}$.
	
We now discuss the rationale behind introducing an approximate transformation rather than the formulation presented in~\eqref{eq:NDAE_to_ODE}. 
We note here that the notion of transforming the NDAE system into a nonlinear ODE model is for reasons beyond numerical simulation and solvability. Howbeit a plethora of numerical methods have been developed to solve DAE systems particular DAEs of index-1~\cite{Abdelmoula2016}. Herein, we are concerned with the aspect of observability for descriptor systems. In~\cite{Sato2013} the concept of algebraic observability is introduced, which formed a local observability definition for DAEs. The study related algebraic observability and local observability through a concept of regulating trajectory. This requires linearizing the NDAE system and writing it in an equivalent ODE system. Another study~\cite{Hou2003} tackled observer design within descriptor systems and formulated the concept of observability using Lie derivatives, however this was validated on a small scale system and was considered to be mathematically limited. 
	
Granted that there is no conventional method for studying observability of NDAEs, we model the descriptor system \eqref{eq:semi_NDAE_rep} by representing the dynamics in an approximate ODE formulation ($\mu$-NDAE). Rather than using the implicit function theorem to represent the power system as~\eqref{eq:NDAE_to_ODE}, we replace the zero on the left-hand side of~\eqref{X_a} by $\mu \dot{\m{x}}_{a}$. The constant $\mu$ is a relatively small number that simulates the system's dynamics while satisfying the power flow constraint equations. The value of $\mu$ is chosen so that a negligible error between the two representations~\eqref{eq:semi_NDAE_rep} and~\eqref{eq:mu-BDF_implicit} is ensured, while also maintaining numerical stability. The rationale behind utilizing such simplistic alternative as compared with~\eqref{eq:NDAE_to_ODE} is evident at the discrete-time modeling level; it offers an alternative to dealing with the existence of partial derivative $\mG_{\m{x}_a}$ and its inverse.
	
The small positive scalar $\mu$ reduces the stiffness of the DAE model, i.e, it transforms the null timescale constant of the algebraic constraints to become of order $\mu$. The value of $\mu$ defines the timescale of the algebraic constraints that are considered as dynamical equations. Therefore, the stability of the $\mu$-NDAE is directly related to the value of $\mu$ chosen and does not depend on algebraic constraints $\m{g}(\cdot)$. This means that the error that bounds the $\mu$-NDAE is directly proportional to the value of $\mu$.

With the proposed approximation, the system is represented as an ODE, albeit without formulating unnecessary computations. The plausibility of such approximation is viable given the low index of the power system model. The validity of such approximation is presented in Section~\ref{sec:case_study} for the proposed discrete-time models.
	
The discrete-time power system in~\eqref{eq:BDF-discrete_NDEA_semi} can be rewritten implicitly for a $\mu$-NDAE as~\eqref{eq:mu-BDF_implicit}, that is denoted by $\m{\phi}(\m{z}_{k},\m{x}_{k-s}):= \m{\phi}(\m{z}_{k})$.

\begin{subequations}\label{eq:mu-BDF_implicit}
	\begin{align}
			\m{0} &= \m{x}_{d,k} -\Sigma_{s=1}^{k_g}\alpha_s \m {x}_{d,k-s} - \tilde{h} \m{f}(\m{z}_{k}),\label{mu-BDF_implicit_1} \\
			\m{0} & = \mu \m{x}_{a,k} -\mu\Sigma_{s=1}^{k_g}\alpha_s \m {x}_{a,k-s} - \tilde{h} \m{g}(\m{x}_{k}).\label{mu-BDF_implicit_2}
	\end{align} 
\end{subequations}
\subsection{NDAE Numerical Solvability: Newton-Raphson Method}\label{sec:NR-Alg}
The solvability of the discretized system in~\eqref{eq:mu-BDF_implicit} involves finding a solution to a set of implicit nonlinear equations, that is, finding $\m{x}_{d}$ and $\m{x}_{a}$ for each time-step $k$. The NR method~\cite{Milano2016,Milano2022} is implemented at each time-step to solve the set of equations under iteration index $i$. The method is iterated until a relatively small error on the $\mathcal{L}_2$--norm of the iteration increment is achieved.
	
First, we rewrite~\eqref{eq:mu-BDF_implicit} as~\eqref{eq:mu-BDF_implicit_discre}, which depicts the convergence of NR's method by introducing iteration index $i$ to the states of the system.
We denote~\eqref{eq:mu-BDF_implicit_discre} as $\m{\phi}(\m{z}^{(i)}_{k},\m{x}_{k-s}) :=\m{\phi}(\m{z}^{(i)}_{k})$, where $\m{z}^{(i)}_{k} :=[\m{x}^{(i)}_{d,k},\m{x}^{(i)}_{a,k},\m{u}^{(i)}_{k}]$ thereby retaining the same definition as $\m{z}_{k}$, however now it is under the NR iteration index $(i)$;

\begin{subequations}\label{eq:mu-BDF_implicit_discre}
	\begin{align}
			\m{0} &=\m{x}^{(i)}_{d,k} -\Sigma_{s=1}^{k_g}\alpha_s \m {x}_{d,k-s} - \tilde{h}\m{f}(\m{z}^{(i)}_{k}),\label{mu-BDF_implicit_discre_1} \\
			\m{0}  & =\mu \m{x}^{(i)}_{a,k} -\mu \Sigma_{s=1}^{k_g}\alpha_s \m {x}_{a,k-s} - \tilde{h}\m{g}(\m{x}^{(i)}_{k}).\label{mu-BDF_implicit_discre_2}
		\end{align} 
\end{subequations}
	
To ensure solution convergence for each time-step $k$, the Jacobian of the nonlinear dynamics in~\eqref{eq:mu-BDF_implicit_discre} is evaluated. Such that, at each time-step $k$ the increment $\Delta \m{x}_{k}^{(i)}$, which is a function of the Jacobian, is evaluated. This increment is then used to update state variable $\m{x}^{(i+1)}_{k}= \m{x}_{k}^{(i)} + \Delta \m{x}_{k}^{(i)}$ for each NR iteration $i$ until the convergence criterion is satisfied. Once NR method converges, time-step $k$ advances to $k+1$ up until the dynamic system is simulated over time-span $t$. The iteration increment $\Delta \m{x}^{(i)}_{k}$ can be written as

\begin{equation}\label{eq:mu-newton_raph}
		\Delta \m{x}^{(i)}_{k} = \left[\mA_{g}(\m{z}^{(i)}_{k})\right]^{-1}\begin{bmatrix}
			\m{\phi}(\m{z}^{(i)}_{k})
		\end{bmatrix},
\end{equation}
where the Jacobian $\m{A}_{g}(\m{z}^{(i)}_{k}) := \begin{bmatrix}
		\tfrac{\partial \m{\phi}(\m{z}^{(i)}_{k})}{\partial \m{x}}
	\end{bmatrix} \in \mathbb{R}^{n\times n}$ is defined as 
\begin{equation}\label{eq:mu-Jac_Newton_Raph}
	\hspace{-0.3cm}	\m{A}_{g}(\m{z}^{(i)}_{k})=
		\begin{bmatrix}
			\eye_{n_d}-\tilde{h} \m{F}_{\m{x}_{d}}(\m{z}^{(i)}_{k}) & -\tilde{h} \m{F}_{\m{x}_{a}}(\m{z}^{(i)}_{k}) \\
			-\tilde{h}\m{G}_{\m{x}_{d}}(\m{x}^{(i)}_{k}) & \mu \eye_{n_a} -\tilde{h} \m{G}_{\m{x}_{a}}(\m{x}^{(i)}_{k})
		\end{bmatrix}.
\end{equation}
	
We define ${n_d} := 4G$ as the number of differential states, $n_a := 2G+2N$ as the number of algebraic states, $n := n_{d}+n_{a}$ as the number of differential and algebraic states. The matrix $\m{F}_{\m{x}_{d}} \in \mathbb{R}^{n_{d}\times n_{d}}$ represents the Jacobian of~\eqref{mu-BDF_implicit_discre_1} with respect to state variable $\m{x}_{d}$, matrix $\m{F}_{\m{x}_{a}}$ $\in \mathbb{R}^{n_{d}\times n_{a}}$ represents the Jacobian of~\eqref{mu-BDF_implicit_discre_1}  with respect to algebraic variables $\m{x}_{a}$, matrix $\m{G}_{\m{x}_{d}}$ $\in \mathbb{R}^{n_{a} \times n_{d}}$ represents the Jacobian of~\eqref{mu-BDF_implicit_discre_2} with respect to state variables $\m{x}_{d}$ and matrix $\m{G}_{\m{x}_{a}}$ $\in \mathbb{R}^{n_{a} \times n_{d}}$ represents the Jacobian of~\eqref{mu-BDF_implicit_discre_2} with respect to algebraic variables $\m{x}_{a}$. Matrix $\eye_{n_d}$ is an identity matrix of dimension similar to $\m{F}_{\m{x}_{d}}$  and $\eye_{n_a}$ is an identity matrix of size similar to $\m{G}_{\m{x}_{a}}$.
	
The discrete-time models under BE and TI discretization methods are presented in Appendix~\ref{apdx:Imp_Disc}. The methodology for solving the NDAE representation of the power system that will be later used to validate the $\mu$-NDAE is presented in Appendix~\ref{apdx:NDAE-disc}.
\section{Initial State Estimation: A MHE Approach}\label{sec:Initial-State-Estimation}
In this section, we develop the MHE framework that is the basis of proposed OPP problem. Based on the discretized time-models developed in Section~\ref{sec:discretenonlinearmodel}, the discrete-time power system dynamics with measurements can be represented as

\begin{subequations}\label{eq:disc_ssm_NDAE}
	\begin{align}
		\hspace{-0.25cm}{\mE_{\mu}}{\m{x}_{k}}&= 
		\begin{cases}
			{\mE_{\mu}}{\m{x}_{k-1}}+ \tilde{h}\eye_{n}
			\begin{bmatrix}
				\m{f}(\m{z}_{k})
				\\ \m{g}(\m{x}_{k})
			\end{bmatrix} &\text{for BE,}\\
			{\mE_{\mu}}{\sum_{s=1}^{k_g}\alpha_s \m {x}_{k-s}} + \tilde{h}\eye_{n}
			\begin{bmatrix}
				\m{f}(\m{z}_{k})
				\\ \m{g}(\m{x}_{k})
			\end{bmatrix} & \text{for BDF,}\\
			{\mE_{\mu}}{\m{x}_{k-1}} + \tilde{h}\eye_{n}
			\begin{bmatrix}
				\m{f}(\m{z}_{k})
				+ \m{f}(\m{z}_{k-1}) \\ \m{g}(\m{x}_{k})+ \m{g}(\m{x}_{k-1})
			\end{bmatrix} & \text{for TI,}
		\end{cases} \label{disc_ss_NDAE} \\
		\m{y}_{k} & = {\m{\tilde{C}}}\m{x}_{k}+ {\m{\tilde{C}}}\m{v}_{k},\label{disc_m_NDAE}
	\end{align} 
\end{subequations}
where matrix $\mE_{\mu} \in \mathbb{R}^{n \times n}$ is a diagonal matrix that has ones on its diagonal for $\m x_d$ and $\mu$ for $\m x_a$. 
%
Diagonal matrix $\m{\Gamma} \in \mathbb{R}^{n_p \times n_p}$ defines the placement of PMUs within the network such that $\m{\Gamma} := \mr{diag}(\m \gamma)$ with PMU mapping vector $\m \gamma:= \{\gamma_i\}_{i=1}^{n_p} \in \{0,1\}^{p}$. Variable $\gamma_{i} = 1$, if a PMU bus is selected and $\gamma_{i} = 0$, otherwise. Under such measurement model, we define $\mathcal{N}_{p}  \subseteq \mathcal{N}$ as the set of buses at which PMUs can be installed, such that $|\mathcal{N}_{p}| := N_p$. We emphasize that the power system is modeled to include both generator and non-generator buses. Consequently, the set $\mathcal{N}_p$ is equivalent to the total number of buses in set $\mathcal{N}.$
The matrix ${\tilde{\mC}} := \m{\Gamma}{\mC} \in \mathbb{R}^{n_p \times n}$ represents the mapping of states variables under the selected PMU configuration, where $n_{p}:= 2{N}_{p}$ represents the number of measured states. 
For the measurement model herein, ${{\mC}}$ is the full measurement mapping matrix that measures the algebraic states $[\m v^{\top} \; \m \theta^{\top}]^{\top}$. Variable $p \leq {N}_{p}$ denotes the number of selected PMUs within the transmission network and $\m{v}_{k}\in \mathbb{R}^{n_{p}}$ is the measurement noise. Discretization constant $\tilde{h}$ for BE and TI discretization methods is defined in Appendix~\ref{apdx:Imp_Disc}.
	
Considering the discretized state-space measurement model with PMU placement presented in~\eqref{eq:disc_ssm_NDAE}, we now introduce the MHE framework under which the observability-based OPP is postulated. The OPP program under the MHE framework utilized is based on the concept of observability for stiff nonlinear networks developed in~\cite{Haber2018}. The rationale behind referring to this approach is that it: $(i)$ adopts a simple open-loop MHE formulation; $(ii)$ allows for the study of observation horizon's influence on state estimation accuracy; $(iii)$ is intrinsically robust against measurement noise~\cite{Alessandri2008}; and $(iv)$ as compared with empirical observability Gramian and other approaches mentioned in Section~\ref{sec:Introduction}, this method---as argued by~\cite{Haber2018}---is the most scalable approach for sensor selection within stiff nonlinear networks. Such approach has also been investigated on traffic networks applications, refer to~\cite{Nugroho2021a}.
	
To that end, we develop the observability analysis through a MHE approach. To begin, we define an observation window, denoted as $\mr{N}_o$, with $\mr{N}_{o}$ discrete measurements. Then, we introduce a nonlinear vector function of the initial state denoted as $\m{h}{(\m \Gamma,\m{x}_0)} := \m{h}{(\m{x}_0)}: \mathbb{R}^{n_p}\times\mathbb{R}^{n} \rightarrow \mathbb{R}^{\mr{N}_o n_p}$. The objective is to minimize the nonlinear least-square error on $\m{h}(\cdot)$ which is posed as \textbf{P1}:
\begin{subequations}\label{eq:MHE_nonlinsq}
		\begin{align}
			(\textbf{P1}) \quad \minimize_{\m{x}_0} \;\;\; &||\m{h}(\m{x}_0)||_2^2 \\
			\text{subject to}  \;\;\;  &  \underline{\m{x}}_0\leq \m{x}_0\leq \overline{\m{x}}_0,
		\end{align}
\end{subequations}
where $\underline{\m{x}}_0$ and $\overline{\m{x}}_0$ are the lower and upper bounds on initial state variables. For power systems, the upper and lower bounds on algebraic variables are obtained from MATPOWER~\cite{Zimmerman2011}.  The vector function $\m{h}{(\cdot)}$ represented in \eqref{eq:g(x_0)} is defined as 
\begin{equation}\label{eq:h(.)}
	\m{h}{(\m{x}_0)} :=\m{y}{(\m{x}_0)} - \m{w}(\m \Gamma,\m x_0),
\end{equation}
where the set of observations over $\mr{N}_{o}$ of the discretized $\mu$-NDAE is represented by vector $\m y(\m x_0):= \{\m{y}_k\}^{\mr{N}_o-1}_{k=1} \in \mathbb{R}^{\mr{N}_o n_p}$. {We note that measurement noise $\m{v}_k$ is considered in~\eqref{eq:g(x_0)} as a result of~\eqref{disc_m_NDAE}, where ~\eqref{disc_m_NDAE} includes the noise vector $\m{v}_k$.
The nonlinear mapping vector function of the dynamics and algebraic states is represented by $\m{w}(\m \Gamma,\m x_0):= \m{w}(\m{x}_{0}) = \{{\tilde{\mC}}\m{x}_k\}^{\mr{N}_o-1}_{k=1} : \mathbb{R}^{n_p}\times \mathbb{R}^{n} \rightarrow \mathbb{R}^{\mr{N}_{o}n_{p}}$. Note that, $\m{w}(\m{x}_0)$ depends on diagonal binary matrix $\m{\Gamma}$ and the system states as defined in~\eqref{disc_ss_NDAE}. As such, vector function $\m{h}{(\cdot)}$ can be written as
\begin{equation}\label{eq:g(x_0)}
	\m{h}{(\m{x}_0)} = \{\m{y}_k\}^{\mr{N}_o-1}_{k=1} - \{{\tilde{\mC}}\m{x}_k\}^{\mr{N}_o-1}_{k=1}.
\end{equation}
\begin{myrem}\label{rem:g(.)}
	The vector $\m h(\cdot)$ is in fact a function of initial state $\m{x}_0$, since the $k$-th state $\m x_k$---as can be observed from \eqref{disc_m_NDAE}---is coupled to $\m x_0$ through the postulated discrete state-space representation.
\end{myrem}
Indeed, for every initial condition $\m{x}_0$, it holds true that $\m{h}{(\m{x}_0)} = 0$ such that, $\m{y}{(\m{x}_0)} = \m{w}(\m x_0)$. We can define the observability of a system with respect to the selected PMU buses according to Definition~\ref{def:inde(2)}.
\begin{mydef}[\cite{Hanba2009} ]\label{def:inde(2)}
	Uniform observability of system \eqref{eq:disc_ssm_NDAE} under the prescribed PMU placement holds true, if for all $\text{inputs} \; \m{u}_k$ there exists a finite measurement horizon $\mr{N}_{o}$ such that the mapping function $\m h(\cdot)$ defined in~\eqref{eq:g(x_0)} is injective, i.e., one-to-one, with respect to initial state $\m{x}_0$.
\end{mydef}
That being said, we can say that for the system to be observable, initial state $\m{x}_0$ under a selected sensor placement has to be uniquely determined for a set of measurements $\m y(\m x_0)$ over horizon $\mr{N}_{o}$. A sufficient condition for $\m h(\cdot)$ to be injective with respect to initial state $\m{x}_0$, is that the Jacobian of $\m h(\cdot)$ around $\m{x}_0$ is full rank~\cite{Hanba2009}.
\subsection{Gauss-Newton for MHE}\label{sec:Gauss}
Under such conditions, we can now solve the nonlinear least squares objective function \eqref{eq:MHE_nonlinsq} by exploiting the discrete nature of the system. We approach solving the least-squares optimization problem numerically using the Gauss-Newton (GN) algorithm. This algorithm has been demonstrated on power systems for DSE~\cite{Minot2016, Saleh2017}. The reasons for referring to a numerical approach rather than utilizing an already developed least-squares solvers are two-fold. The first is that GN algorithm is more computationally efficient and leads to solution convergence faster. The second is that under the latter existing solvers approach, in particular while considering large networks, i.e, ACTIVSg200-bus case, MATLAB's \textit{lsqminorm} solver could not converge to an initial state estimate.
	
To solve \textbf{P1} using GN, we reformulate the objective and pose it as the minimization of the $\mathcal{L}_{2}$--norm of the residual function vector $\m{r}(\m \Gamma,\m{q})$. The residual concatenates $(i)$ the measurement equation~\eqref{disc_m_NDAE} and $(ii)$ the discretized $\mu$-NDAE model~\eqref{disc_ss_NDAE}. The redefined optimization problem \textbf{P1} is posed as \textbf{P2}:
\begin{equation}\label{eq:GaussN_nonlinsq}
		(\textbf{P2}) \quad \minimize_{\m{q}_0} \quad ||\m{r}(\m \Gamma,\m{q})||_2^2,
\end{equation}
where the vector $\m{q}  \in \mathbb{R}^{\mr{N}_{o}n}$ concatenates the dynamic and algebraic states simulated over horizon $\mr{N}_o$. This can be written as $\m{q}  := \{\m{q}_{k}\}^{\mr{N}_{o}-1}_{k=0}  = [\m{x}^{\top}_{d,0}, \; \m{x}^{\top}_{a,0},\; \dots, \;\m{x}^{\top}_{d,\mr{N}_{o} -1},  \; \m{x}^{\top}_{a,\mr{N}_{o} -1} ]^{\top}$. As such, the residual vector $\m{r}(\m \Gamma,\m{q}):= \m{r}(\m{q}) \in \mathbb{R}^{\mr{N}_{o} n_p+\mr{N}_{o}n}$ is as follows
\begin{equation}\label{eq:residual}
		\m{r}(\m{q}) := \begin{bmatrix}
			\m{r}_{\m{y}}^{{\top}} & \m{r}_{\m{x}}^{{\top}}
		\end{bmatrix}^{\top},
\end{equation}
where vector $\m{r}_{\m{y}} := \m{h}{(\m{x}_0)} = \{\m{r}_{\m{y}_k}\}^{\mr{N}_{o}-1}_{k=0} \in \mathbb{R}^{\mr{N}_{o} n_p}$ is the residual function of the measurement equation for $\mr{N}_o$ observations that is defined as \eqref{eq:ry}, such that $\hat{\m{x}}_{k} \in \mathbb{R}^{\mr{N}_{o} n}$ is the vector representing the estimated differential and algebraic states
\begin{equation}\label{eq:ry}
		\m{r}_{\m{y}} := \m{y}_{k} - \tilde{\m C}\hat{\m{x}}_{k},
\end{equation}
and vector $\m{r}_{\m{x}} :=\m{\phi}(\m x_0) =\{\m{r}_{\m{x}_k}\}^{\mr{N}_{o}-1}_{k=0} \in \mathbb{R}^{\mr{N}_{o} n}$ is the residual of the discretized $\mu$-NDAE model, where $\m{r}_{\m{x}_{k}} $ for time-step $k$ is defined as

\begin{equation}\label{eq:rx}
		\hspace{-0.25cm}\m{r}_{\m{x}_{k}}\hspace{-0.1cm} :=\hspace{-0.1cm} \begin{cases}
			{\mE_{\mu}}(\hat{\m{x}}_{k} - \hat{\m{x}}_{k-1}) - \tilde{h}\eye_{n}
			\hspace{-0.1cm}\begin{bmatrix}
				\m{f}(\hat{\m{x}}_{k})
				\\ \m{g}(\hat{\m{x}}_{k})
			\end{bmatrix}&\hspace{-0.3cm} \text{for BE,}\\
			{\mE_{\mu}}({\hat{\m{x}}_{k}}-{\sum_{s=1}^{k_g}\alpha_s \hat{\m{x}}_{k-s}}) - \tilde{h}\eye_{n}
			\hspace{-0.1cm}\begin{bmatrix}
				\m{f}(\hat{\m{x}}_{k})
				\\ \m{g}(\hat{\m{x}}_{k})
			\end{bmatrix} & \hspace{-0.3cm}\text{for BDF,}\\
			{\mE_{\mu}}(\hat{\m{x}}_{k} -\hat{\m{x}}_{k-1})-\tilde{h}\eye_{n}
			\hspace{-0.1cm}	\begin{bmatrix}
				\m{f}(\hat{\m{x}}_{k})
				+ \m{f}(\hat{\m{x}}_{k-1}) \\ \m{g}(\hat{\m{x}}_{k})+ \m{g}(\hat{\m{x}}_{k})
			\end{bmatrix} &\hspace{-0.3cm}\text{for TI.}
		\end{cases}
\end{equation}

Having developed the residual function that is the objective of \textbf{P2}, we move forward with solving the minimization problem using the GN iterative method by updating state vector $\m{q}$ such that \eqref{eq:GaussN_nonlinsq} is minimized. The GN update for iteration $i$ is given as~\eqref{eq:Gauss-itr} with a GN step-size denoted by $h_g$.
\begin{equation}\label{eq:Gauss-itr}
		\hspace{-0.3cm}\m{q}^{(i+1)} = \m{q}^{(i)} -
		h_g\big(\m{J}_g(\m{q}^{(i)})^{\top}\m{J}_g(\m{q}^{(i)})\big)^{-1}\m{J}_g(\m{q}^{(i)})^{\top} \m{r}(\m{q}^{(i)}),\hspace{-0.1cm}
\end{equation}
where Jacobian matrix in~\eqref{eq:Gauss-itr} of the residual function $\m{r}(\m{q})$ is
\begin{equation}\label{eq:Jac_Gauss}
		\m{J}_g(\m \Gamma, \m{q}^{(i)}) := \m{J}_g(\m{q}^{(i)}) =  \begin{bmatrix}
			\m M \\ \m N
		\end{bmatrix},
\end{equation}
where Jacobian matrix of residual function $\m{r}_{\m y}$ is denoted by $\m M$ and defined as $\m M := \mr{blkdiag}(-\tilde{\m C}) \in \mathbb{R}^{\mr{N}_{o} n_p \times \mr{N}_{o} n}$. The Jacobian matrix of residual function $\m{r}_{\m x}$ is denoted by $\m N$ and defined as $\m{N}:=\mr{blkdiag}(\m{A}_g)\in \mathbb{R}^{\mr{N}_{o} n \times \mr{N}_{o} n}$. Herein, $\m A_g \in \mathbb{R}^{n\times n}$ is the Jacobian of the discretized $\mu$-NDAE~\eqref{disc_ss_NDAE} which is evaluated for observation horizon $\mr{N}_{o}$ and is therefore dependent on the discretization method. Such that, for the BE and BDF discretization methods, the Jacobian matrix $\m A_g$ is defined as in~\eqref{eq:mu-Jac_Newton_Raph}, and for the TI method, $\m A_g$ is in Appendix~\ref{apdx:Imp_Disc}. With the iteration update defined, Gauss-Newton iterative method is performed until the $\mathcal{L}_2$--norm of the residual \eqref{eq:residual} is minimized. Algorithm~\ref{alg:Disc-Alg} in Appendix~\ref{apdx:alg_MHE} outlines the proposed MHE for initial state estimation using the GN method.
\section{Observability-based OPP in Power Networks}\label{sec:OSP-Power-system}
In this section, we formulate the observability-based OPP that is based on the discretized system dynamics and MHE framework developed in Sections~\ref{sec:discretenonlinearmodel} and \ref{sec:Initial-State-Estimation}. 

To quantify observability of the power system, the concept of observability through the observability Gramian is used. Observability metrics that allow us to numerically quantify observability taking into account different aspects of the observability Gramian include: the condition number, rank, smallest eigenvalue, trace and determinant. Interested readers are referred to~\cite{Qi2015, Summers2016}, both of which present a more elaborate discussion on the different metrics that quantify observability of dynamical systems. For the OPP within the scope of this work, the trace of the observability Gramian is considered. The trace quantifies the average observability in all directions of the state-space. The determinant is usually also considered since it is able to measure observability in the noise-space. However, given the MHE approach that the placement formulation is built upon, redundancy towards noise is already considered prior to building the observability matrix.
\begin{myrem}\label{rem:Condition}
	Additional consideration should be given if the observability Gramian has a large condition number. This implies that observability is ill-conditioned and that any perturbation to initial state $\m{x}_0$ changes the observability significantly. 
\end{myrem}

With that in mind, we pose the OPP problem as the trace maximization of the observability matrix, while checking for near-zero eigenvalues. We implement the OPP on standard optimization interfaces such as YALMIP~\cite{Lofberg2004} along with the Gurobi~\cite{gurobi} solver. The OPP problem for the discretized state-space measurement model~\eqref{eq:disc_ssm_NDAE} can be defined for a fixed number of PMUs as
\begin{subequations}\label{eq:ssp_gramian_trace}
	\begin{align}
		\hspace{-0.5cm}(\textbf{P3})\;\; \minimize_{\m{\Gamma}} \;\;\;&
		-\mathrm{trace}\left(\m W_o(\m{\Gamma},{\m x}_0)\right)\vspace{-0.5cm}\label{eq:ssp_gramian_trace_1}\vspace{-1cm}\\
		\text{subject to} \;\;\;  & \sum^n_{i=1} \gamma_i  = p, \;\;\;
		\gamma_i \in \{0,1\},\label{eq:ssp_gramian_trace_2}
	\end{align}
\end{subequations} 
where $\mW_o(\cdot) \in \mathbb{R}^{n\times n}$ is the observability matrix of the discretized $\mu$-NDAE system. The observability matrix for a nonlinear descriptor system under a MHE formulation with OPP can be written as
\begin{equation}\label{eq:obs_gram}
	\m W_o(\m{\Gamma},{\m x}_0) = \m{J}^T(\m{\Gamma},{\m x}_0)\m{J}(\m{\Gamma},{\m x}_0),
\end{equation}
where the matrix $\m J(\cdot) \in \mathbb{R}^{\mr{N}_{o}n_p \times n}$ represents the Jacobian of~\eqref{eq:g(x_0)} around $\m{x}_0$ for horizon $\mr{N}_o$ and can be defined as
\begin{equation}\label{eq:Jac_obs_horz}
		\m J(\m \Gamma,\m{x}_0) := \begin{bmatrix}
			\eye_{\mr{N}_o} \kron \tilde{\m C}
		\end{bmatrix}
		\begin{bmatrix}
			\eye_n \\ \frac{\partial \m{x}_1}{\partial \m{x}_0} \\ 
			\vdots \\
			\frac{\partial \m{x}_{\mr{N}_{o}-1}}{\partial \m{x}_0}\\ 
		\end{bmatrix},
\end{equation}
where the identity matrix $\eye_{\mr{N}_o} \in \mathbb{R}^{\mr{N}_o\times\mr{N}_o}$ is of size equal to the observation horizon. The partial derivatives in~\eqref{eq:Jac_obs_horz} can be expressed as
\begin{equation}\label{eq:partial_jac}
		\tfrac{\partial \m{x}_{j+1}}{\partial \m{x}_{j}} =\tfrac{\partial \m{x}_{j+1}}{\partial \m{x}_{j}}|_{\m{x}_j} \quad \text{for }j =\{  0, \;1\;, \dots\;, \mr{N}_{o} -1\}.
\end{equation} \vspace{0.2cm}

The observability formulation herein maps the OPP problem under a quantitative measure of DSE. That being said, the mapping of sensor location is represented by the matrix ${\tilde{\m C}}$, where under full PMU placement---that is, when all states are measured---the observability Gramian $\m W_o(\m{\Gamma},{\m x}_0)$ is maximum. Conversely, under zero sensing, i.e., ${\tilde{\m C}} $ is equal to a zero matrix, a maximum state estimation error is achieved.

As mentioned in Section~\ref{sec:Initial-State-Estimation}, the rank of the Jacobian in~\eqref{eq:Jac_obs_horz} is $\rank(\m{J}(\m\Gamma,\m{x}_0)) = n_d + n_a =n \; \forall \; \m {x}_0$. Note that, this holds true for the power system cases considered in the numerical studies section, thereby implying uniform observability. As for calculating the Jacobian in~\eqref{eq:Jac_obs_horz}, knowledge of $ \m x_k \; \forall \; j = \{0\;,1\;,\dots\;,\mr{N}_{o}-1\}$ is required. This can be obtained by simulating the discrete-time $\mu$-NDAE dynamics over $\mr{N}_{o}$. To calculate the partial derivatives terms in~\eqref{eq:Jac_obs_horz}, we apply the chain rule to compute the $j$-th partial derivative as follows $\tfrac{\partial \m{x}_{j}}{\partial \m{x}_0} = \tfrac{\partial \m{x}_j}{\partial \m{x}_{j-1}} \cdots \tfrac{\partial \m{x}_{1}}{\partial \m{x}_0}$. However, given the implicit nature of the discretized $\mu$-NDAE system, the computation of $\tfrac{\partial \m{x}_{j}}{\partial \m{x}_0} $ for the power system is not straightforward and depends on the discretization method chosen~\cite{Haber2018}. For Gear's method, the $j$-th partial derivative is computed as follows $\frac{\partial \m{x}_{j}}{\partial \m{x}_0} = \frac{\partial \m{x}_j}{\partial \m{x}_{j-{k}_{g}}}\cdots \frac{\partial \m{x}_{{k}_{g}}}{\partial \m{x}_0}$. The rationale behind changing the chain rule expression for the BDF method is described in Appendix~\ref{apdx:Jac_Gear}.

Moreover, computing $\m J(\cdot)$ for NDAEs is non-trivial. This is due to calculating the partial derivative of algebraic states, where obtaining an explicit representation of the partial derivative $\frac{\partial \m{x}_{a,j+1}}{\partial \m{x}_{a,j}}$ for the algebraic states is non-trivial unless the system dynamics is reformulated into an ODE representation~\eqref{eq:NDAE_to_ODE}. Additional information regarding this complexity is provided in Appendix~\ref{apdx:calc_Jac}.
We present the Jacobian for the different discrete-time models \eqref{disc_ss_NDAE} that originate from the implicit state-space equations in Appendix~\ref{apdx:calc_Jac}. Specifically, we express the partial derivative $\frac{\partial \m{x}_{j+1}}{\partial \m{x}_{j}}$  in explicit form for each of the discrete-time models under study.
\\

Having formulated the observability Gramian we now discuss reformulating the OPP problem \textbf{P3}. One approach for tackling the combinatorial class of sensor selection problems within networks, is posing such problem as a set function optimization problem. Such that for a submodular\footnote[1]{A function $ \mathcal{F} : 2^V \rightarrow \mathbb{R} $ is \emph{submodular}  if for every $A,B \subseteq V$, and $e \in V \backslash B$ it holds that $\Delta(e|A) \geq \Delta(e|B)$. Equivalently, a function $ \mathcal{F} : 2^V \rightarrow \mathbb{R} $ is \emph{submodular} if for every $A,B \subseteq V$ it holds that $\mathcal{F}(A\cap B) +\mathcal{F}(A\cup B) \leq \mathcal{F}(A) +\mathcal{F}(B)$.} objective function, solving a set maximization problem is a common approach. Intrinsically, submodularity is considered to be a diminishing returns property~\cite{Lovasz1983}. Accordingly, the OPP in \textbf{P3} can be posed as a set function optimization program denoted by \textbf{P4}. The set of selected sensors is denoted by $\mathcal{{\m{Z}}}\subseteq \mathcal{N}_p$. The mapping of selected PMUs in set $\mathcal{\m{Z}}$ is encoded by the parameterized measurement mapping matrix ${\tilde{\m C}}$. 
\begin{subequations}\label{eq:OSP_Modular}
	\begin{align}
		(\textbf{P4})\;\; \minimize_\mathcal{{\m{Z}}} \;\;\;&
		-\mathrm{trace}\left(\m{W}_o(\mathcal{{\m{Z}}} ,{\m x}_0)\right)\label{eq:OSP_Modular_1}\\
		\text{subject to} \;\;\;  & |\mathcal{{\m{Z}}} | = p, \;\;
		\mathcal{{\m{Z}}}  \subseteq \mathcal{N}_p.\label{eq:OSP_Modular_2}
	\end{align}
\end{subequations}
Submodular set maximization problem is still considered an NP-hard integer program. A common computationally tractable approach that achieves a suboptimal solution for maximizing monotone increasing\footnote[2]{A set function $\mathcal{F} : 2^V \rightarrow \mathbb{R}$ is monotone increasing  if $\forall \; A,B \subseteq V$ the following holds true; $A \subseteq B \rightarrow \mathcal{F}(A) \leq \mathcal{F}(B)$} submodular functions can be performed by a greedy heuristic approach. Solving the OPP under a greedy approach yields suboptimal solutions that are at least $(1-1/e) = 63\%$ of the optimal solution~\cite{Calinescu2016}.

Considering the above, we revisit the OPP posed in \textbf{P4} that is solved as a submodular set optimization program and instead pose it as an \textit{a priori set optimization program}. The idea is based on the \textit{a priori} observability knowledge from individual sensor measurements. The proposed framework involves computing prior singular contribution resulting from each PMU placement on the observability matrix. After saving such \textit{a priori} information regarding observability contributions, the OPP that is then posed as a convex integer program (IP) is solved. The plausibility of such approach stems from the fact that the observability matrix $\m W_o(\cdot)$ is a modular\footnote[3]{ A set function is modular if it is both submodular and supermodular, such that $\forall \; A,B \subseteq V$ the following holds true: $	\mathcal{F}(A\cap B) +\mathcal{F}(A\cup B) = \mathcal{F}(A) +\mathcal{F}(B)$. Supermodularity of a set function $\mathcal{F}(\cdot)$ holds true if $-\mathcal{F}(\cdot)$ is submodular.} set function. In the context of linear systems,~\cite{Summers2016} showed that the observability matrix retains a modular set function structural property. With regards to the nonlinearities of model under study in this work, we prove that the observability matrix under PMU placement is modular with respect to decision variable $\m{\Gamma}$. 
The idea of considering the modularity of the observability Gramian is that a modular function forms positive linear combinations of the single elements in the modular set. This intuitively can be explained in the sense that modular functions and linear functions are analogous such that, each element that belongs to the set that forms the modular function has an independent contribution to the function value. With that in mind, the next proposition formulates the observability matrix $\m W_o(\cdot)$ as a linear combination of its individual elements. 
\begin{myprs}\label{prop:modularity}
	The observability matrix $\m W_o(\cdot)$ can be written as a linear combination of individual observability matrices computed from each individual PMU contribution as follows 
		$$\m{W}_{o}({\mathcal{\m{Z}}} ,{\m x}_0) = \sum_{i = 1}^{N_{p}} \m{W}_{o,i}(\mathcal{{\m{Z}}}_{i},\m{x}_0),$$
		where $\mathcal{Z}_{i}$ corresponds to the selected i-th sensor that is encoded in matrix ${\tilde{\m C}}$. That is, $\mathcal{Z}_{i}$ is a binary set that has a value of 1 at the i-th selected sensor location and zeros elsewhere, as such $|\mathcal{Z}_{i}| = 1$.
\end{myprs}
Proof of Proposition~\ref{prop:modularity} is included in Appendix~\ref{apdx:proof}. Accordingly, for each element in set $\mathcal{N}_{p}$ we evaluate $\m{W}_{o,i}(\mathcal{{\m{Z}}}_{i},\m{x}_0) \; \forall \; i = \{1,\dots,N_p\}$, prior to solving the OPP problem. 
With that in mind, the~\textit{a priori set optimization program} for optimal PMU placement denoted by~\textbf{P5} can be posed as
\begin{subequations}\label{eq:apriori_OSP_Modular}
	\begin{align}
		(\textbf{P5})\;\; \minimize_\mathcal{{\m{Z}}} \;\;\;&
		-\mathrm{trace}\left(\m W^{1}_o(\mathcal{{\m{Z}}} ,{\m x}_0)\right)\label{eq:apriori_OSP_Modular_1}\\
		\text{subject to} \;\;\;  & |\mathcal{{\m{Z}}} | = p, \;
		\mathcal{{\m{Z}}}  \subseteq \mathcal{N}_p,\label{eq:apriori_OSP_Modular_2}
	\end{align}
\end{subequations}
where $\m W^{1}_o(\mathcal{{\m{Z}}} ,{\m x}_0) = \sum_{i = 1}^{N_{p}} \m W_{o,i}(\mathcal{Z}_{i},\m{x}_0)$.

The concept of \textit{a priori} optimization has been proposed before in optimization, in particular combinatorial optimization. Maros~\cite{Maros1990} introduced the concept of \textit{a priori} optimization for optimizing  randomly distributed networks in a computationally efficient manner. This concept encompasses attaining individual instance contribution knowledge prior to solving the combinatorial problem whilst not having exponentially complex computations being performed during each optimization instance, i.e, at each instance the complex computations are already evaluated. This allows one to perform combinatorial optimization with minimal computing power. 

\begin{figure}[t!]
	\centering
	\includegraphics[keepaspectratio=true,scale=1]{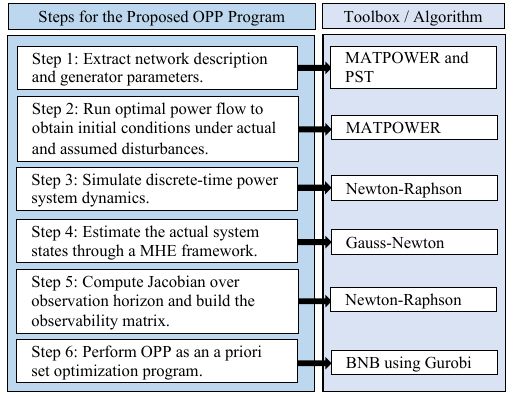}
	\vspace{-0.5cm}
	\caption{Implementation of OPP framework for an NDAE power systems.}\label{frame:NDAE-OSP-Alg}
	\vspace{-0.4cm}
\end{figure}

Having provided \textit{a priori} information for a particular instance which in our case is possible given the modular nature of the observability matrix. Consequently,~\textbf{P5} categorized as a convex integer program (IP), is considered computationally less exhaustive and therefore scalable to large power networks.  

With such formulation,~\textbf{P5} can be solved efficiently given that the observability metric~\eqref{eq:apriori_OSP_Modular_1} is evaluating a linear combination of the individual pre-calculated contributions from each of the selected PMU buses. The implementation of this OPP framework for a complete descriptor power system is summarized in Fig.~\ref{frame:NDAE-OSP-Alg}. The validity and effectiveness of this approach are studied in the subsequent section of this paper.

\section{Case Studies: Validation and Results}\label{sec:case_study}
In this section, we first validate the discrete-time $\mu$-NDAE system developed in Section~\ref{sec:discretenonlinearmodel} and then we evaluate various aspects of the proposed OPP problem~\textbf{P5}. The objective is to obtain an optimal PMU placement for a specified sensor fraction $p$ that yields an observable system under load/renewables uncertainty. As such, we attempt to answer the following questions:
\begin{itemize}[leftmargin=*]
	\item \textbf{Q1}: What is an appropriate value for $\mu$ that offers a good compromise between numerical stability and accuracy in simulating the power system dynamics?
	\item \textbf{Q2}: How does the choice of discretization method affect the power system simulation, and ultimately, the OPP.?
	\item \textbf{Q3}: Are the optimal PMU placements robust against load/renewables uncertainty and measurement noise?
	\item \textbf{Q4}: Does the framework under which we pose the OPP problem result in modular PMU placements and what is the significance of such modularity?
\end{itemize}

The simulations and optimization problem are performed in MATLAB R2021b running on a Macbook Pro having an Apple M1 Pro chip with a 10-core CPU and 16 GB of RAM. The OPP program is interfaced on MATLAB through YALMIP~\cite{Lofberg2004} and implemented using a standard branch and bound method (BNB) with Gurobi~\cite{gurobi} as the solver. 

We consider three different power networks for the assessment of the proposed approach:
\begin{itemize}[leftmargin=*]
	\item $\mr{case}$-$\mr{9}$: Western System Coordinating Council (WSCC) 9-Bus network (9-bus system with 3 synchronous generators).
	\item $\mr{case}$-$\mr{39}$: IEEE 39-Bus network "New-England Power System" (39-bus system with 10 synchronous generators).
	\item $\mr{case}$-$\mr{200}$: ACTIVSg200-Bus network "Illinois200 case" (200-bus system with 49 synchronous generators).
\end{itemize}
\subsection{OPP Implementation}\label{sec:OPP_Frame}
The test cases can be downloaded online from the Illinois center for a smarter electric grid cases repository~\cite{Birchfield2017}.
The generator parameters are extracted from power systems toolbox (PST)~\cite{Sauer2017} case file $\mr{data3m9b.m}$ and $\mr{datane.m}$ for $\mr{case}$-$\mr{9}$ and $\mr{case}$-$\mr{39}$ respectively. For $\mr{case}$-$\mr{200}$ the generator parameters are chosen based on the ranges provided in the PST toolbox. Regulation and chest time constants for the generators are chosen as $R_{\mr{D}i} = 0.2 \; \mr{Hz/pu}$ and $T_{\mr{CH}i} = 0.2 \; \mr{sec}$, since they are not included in the PST case file. The steady-state initial conditions (before disturbance) for the power system are generated from solutions of the power flow obtained from MATPOWER $\mr{runpf}$ function~\cite{Zimmerman2011}. The synchronous speed is set to $\omega_{0} = 120\pi \;  \mr{rad/sec}$ and a power base of $100 \; \mr{MVA}$ is considered for the power system. The aforementioned steps cover the first two steps from the OPP framework illustrated in Fig.~\ref{frame:NDAE-OSP-Alg}.

\begin{table}
	\fontsize{9}{9}\selectfont
	\centering 
	\caption{RMSE value of the system states simulated using Matlab $\mr{ode15i}$ and the different discretization methods.}
	\label{tab:ODE_disc}
	\vspace{-0.15cm}
	\renewcommand{\arraystretch}{1.5}
	\begin{tabular}{l|l|l|l|l}
		\midrule \hline
		\multirow{2}{*}{Network} & \multicolumn{1}{c|}{Disturbance}                             & \multicolumn{3}{c}{$\mr{RMSE}$}                           \\ \cline{2-5} 
		& \multicolumn{1}{c|}{$\alpha_{L}$} & \multicolumn{1}{c|}{BE} & \multicolumn{1}{c|}{BDF} & \multicolumn{1}{c}{TI} \\ \hline
		\multirow{3}{*}{$\mr{case}$-$\mr{9}$} 
		& \multicolumn{1}{c|}{$2\%$} & \multicolumn{1}{c|}{$0.0022$} & \multicolumn{1}{c|}{$1.2857 \times 10^{-5}$} & \multicolumn{1}{c}{$0.0022$} \\ \cline{2-5} 
		& \multicolumn{1}{c|}{$3\%$} & \multicolumn{1}{c|}{$0.0049$} & \multicolumn{1}{c|}{$2.0379  \times 10^{-5}$} & \multicolumn{1}{c}{$0.0048$} \\ \cline{2-5} 
		& \multicolumn{1}{c|}{$4\%$}  & \multicolumn{1}{c|}{$0.0126$} & \multicolumn{1}{c|}{$9.5091  \times 10^{-5}$}& \multicolumn{1}{c}{$0.0122$}  \\ \hline
		\multirow{3}{*}{$\mr{case}$-$\mr{39}$} 
		& \multicolumn{1}{c|}{$3\%$} & \multicolumn{1}{c|}{$0.2109$} & \multicolumn{1}{c|}{$0.0134$} & \multicolumn{1}{c}{$0.1998$}  \\ \cline{2-5} 
		& \multicolumn{1}{c|}{$5\%$} & \multicolumn{1}{c|}{$0.2171$} & \multicolumn{1}{c|}{$0.0139$} & \multicolumn{1}{c}{$0.1908$} \\ \cline{2-5} 
		& \multicolumn{1}{c|}{$7\%$} & \multicolumn{1}{c|}{$0.2418$} & \multicolumn{1}{c|}{$ 0.0172$}& \multicolumn{1}{c}{$0.2053$} \\ \hline 
		\multirow{3}{*}{$\mr{case}$-$\mr{200}$} 
		& \multicolumn{1}{c|}{$10\%$} & \multicolumn{1}{c|}{$0.0129$} & \multicolumn{1}{c|}{$ 1.1396  \times 10^{-5}$} & \multicolumn{1}{c}{$0.0131$}\\ \cline{2-5} 
		& \multicolumn{1}{c|}{$15\%$} & \multicolumn{1}{c|}{$0.0185$} & \multicolumn{1}{c|}{$0.0010$} & \multicolumn{1}{c}{$0.0186$} \\ \cline{2-5} 
		& \multicolumn{1}{c|}{$20\%$}  & \multicolumn{1}{c|}{$0.0227$} & \multicolumn{1}{c|}{$0.0014$} & \multicolumn{1}{c}{$0.0228$} \\ \cline{2-5} 
		\toprule \bottomrule
	\end{tabular}
	\vspace{-0.3cm}
\end{table}

\begin{figure*}[t!]
	\centering 
	\vspace{-0.2cm}
	\subfloat{\includegraphics[keepaspectratio=true,scale=0.7]{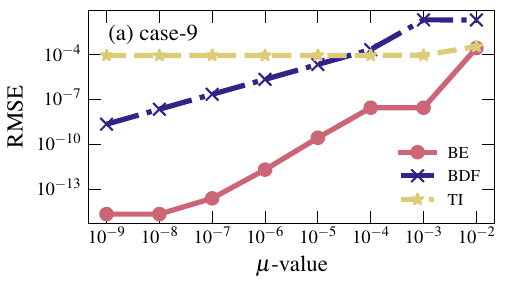}}
	\subfloat{\includegraphics[keepaspectratio=true,scale=0.7]{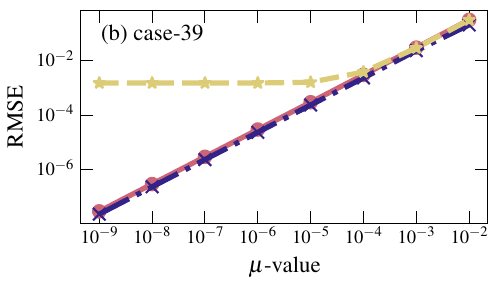}}
	\subfloat{\includegraphics[keepaspectratio=true,scale=0.7]{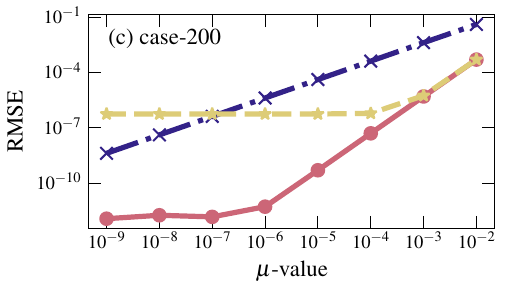}}
	\vspace{-0.25cm}
	\caption{RMSE on both dynamic and algebraic states between the NDAE and $\mu$-NDAE discrete-time representations of the power systems: (a) $\mr{case}$-$\mr{9}$ ($\alpha_{L} = 2\%$), (b) $\mr{case}$-$\mr{39}$ ($\alpha_{L} = 5\%$), and (c) $\mr{case}$-$\mr{200}$ ($\alpha_{L} = 20\%$).}\label{fig:disc-mu}
	\vspace{-0.3cm}
\end{figure*}

To simulate the discretized descriptor system using the NR algorithm (see, Section~\ref{sec:NR-Alg}), we set the discretization step-size $h = 0.1$ and simulation time-span $t = 30 \; \mr{sec}$. Starting from the initial steady-state conditions obtained from solving the power flow equations (Steps 1 and 2), we introduce a load disturbance at $t > 0$ on initial load $(\mr{P}_\mr{L}^0,\mr{Q}_\mr{L}^0)$. In this model renewables are modeled as a negative load, that is renewables are considered to inject power into the network as given in~\eqref{eq:power_balance}. The total power generation considered within the 3 cases is $(\mr{P}_\mr{R}^0,\mr{Q}_\mr{R}^0)=(0.2\mr{P}_\mr{L}^0,0.2\mr{Q}_\mr{L}^0)$. The perturbed magnitude under a load disturbance $(\alpha_{L})$ is computed as $(\mr{\tilde{P}}_\mr{L}^0,\mr{\tilde{Q}}_\mr{L}^0 ) = (1+\tfrac{\alpha_{L}}{100})(\mr{P}_\mr{L}^0,\mr{Q}_\mr{L}^0 )$. Moreover, the perturbed magnitude under a renewable disturbance $(\alpha_{R})$ is computed as $(\mr{\tilde{P}}_\mr{r}^0,\mr{\tilde{Q}}_\mr{r}^0 ) = (1+\tfrac{\alpha_{R}}{100})(\mr{P}_\mr{R}^0,\mr{Q}_\mr{R}^0 )$. Under the scope of this paper, we demonstrate simulating the system dynamics with load disturbance magnitude $\alpha_{L} )$ varying between $\{2 \%, 20\% \}$ of the unperturbed initial loads and with renewable disturbance magnitude $\alpha_{R} = \alpha_{L}$ of the unperturbed initial renewable loads (Step 3, see Fig.~\ref{frame:NDAE-OSP-Alg}). Note that, by inducing transient conditions from the perturbed load/renewables injections, the system's response is automatically regulated by the governor response $\m{T}_\mr{r}$ modeled in~\eqref{eq:differential_dynamics_4}. This means that the input response is different for each of the simulated transient conditions.

System observability analysis is then performed in order to solve the OPP problem posed as~\textbf{P5}, with an aim to seek an optimal configuration of PMU placement represented by set $\mathcal{{\m{Z}^{*}}}$ under a maximum number of PMUs denoted by $p$. To begin, we first initialize a power system under assumed initial conditions ${\bar{\m{x}}}_{0}$ that has been perturbed under load/renewables disturbances with $\alpha_{R} = \alpha_{L}= 4\%$. Then by simulating the discretized measurement model in~\eqref{eq:disc_ssm_NDAE} and under $v = 2\%$ Gaussian measurement noise over observation horizon $\mr{N}_o$, we perform initial state estimation assuming full PMU placement, that is, $|\mathcal{{\m{Z}}}| = n_p$. Note that the Gaussian noise models a random noise vector that follows a normal distribution with a mean of zero and a standard deviation of $\sigma(1+v)$, with $v$ as the noise level. The GN method developed in Section~\ref{sec:Initial-State-Estimation} for the MHE is implemented to solve for initial state estimate $\hat{\m{x}}_0$ under optimization problem~\textbf{P2}. As for the GN algorithm constants, we set time-step constant $h_g = 0.1$ and tolerance on residual as $10^{-4}$. Then, based on the initial state estimate, the observability of the system is computed under the proposed modularity formulation outlined in~Section~\ref{sec:OSP-Power-system}. The observation horizon window is chosen as $\mr{N}_{o} = t/h= 300.$

The OPP problem~\textbf{P5} is then solved to obtain optimal set $\mathcal{{\m{Z}}}^{*}$ and compute the estimation error resulting from the optimal PMU placements. The estimation error that is based on the estimate of the GN algorithm is computed as $\eps := \tfrac{\norm{\hat{\m{x}}_0-\m{x}_0}_2}{\norm{\m{x}_0}_2}$, where $\m {x}_{o}$ is the actual state that we want to estimate and $\hat{\m {x}}_o$ is its estimate computed by solving the nonlinear least squares problem \textbf{P2} for the fixed PMU location $\mathcal{{\m{Z}}}^{*}$. This concludes the all the steps required to implement the OPP framework depicted in Fig.~\ref{frame:NDAE-OSP-Alg}. It is noteworthy to mention that~\textbf{P5} is classified as a convex integer program (IP) since the presumed initial state estimate $\hat{\m{x}}_0$ is fixed and binary vector $\m{\Gamma}$ is the optimization variable.

\subsection{Simulating the Discretized Power System Dynamics}\label{sec:case_simulate}
To assess the accuracy of the discretization methods presented, we first simulate the baseline system dynamics using MATLAB DAE solver $\mr{ode15i}$ under the perturbations mentioned above (see, Step $1-3$). Then, we simulate the dynamics using the discretization methods developed in Section~\ref{sec:discretenonlinearmodel}. Finally, we calculate the root mean square error $(\mr{RMSE})$ for each discretization over time period $t$. This is calculated as $\mr{RMSE} := \sqrt{\tfrac{\sum_{k=1}^{t} \m{e}_{k}^{2}}{t}}$, where $\m{e}_{k} := \abs{\m{\tilde{x}}_{k} - \m{x}_{k}}$ is the difference between the states of the two system representations with $\m{\tilde{x}}_{k}$ corresponding to the discretized system and $\m{x}_{k}$ to the system solved using $\mr{ode15i}$. The setting chosen for $\mr{ode15i}$ are: $(i)$ absolute tolerance as $1\times 10^{-05}$, $(ii)$ relative tolerance as $1\times 10^{-04}$ and $(iii)$ maximum step-size equal to $0.001$. As for the Newton-Raphson algorithm, we set: $(i)$ absolute tolerance on $\mathcal{L}_2$--norm of iteration convergence as $10^{-2}$ and $(ii)$ maximum iterations as $10$.
The results are summarized in Tab.~\ref{tab:ODE_disc}. It can be seen that for each of the three methods and under different load perturbations, the BDF discretization method outperforms BE and TI methods by having the lowest $\mr{RMSE}$ values. We note here that the BDF discretization order chosen for the simulations is $k_g=3$. Also, under increased load/renewable uncertainty the systems results in a larger $\mr{RMSE}$ value on the state trajectories. This increase is expected since the perturbations induce transient conditions within the system. Such transients exhibit stiff nonlinearities and might render the system unstable.

\subsection{Validating the Discrete $\mu$-NDAE Model}\label{sec:case_validate}
We now move forward with validating the approximate $\mu$-NDAE system. The validity of such approach is demonstrated for the three case systems by choosing different values for $\mu$ over a range of $\mu = \{10^{-2}, 10^{-9}\}$. We note that for any $\mu> 10^{-2}$, the system does not converge to a solution; that is, the power balance equations are not satisfied. Thus, numerical instability in the simulation becomes evident for any value greater than $\mu= 10^{-2}$. The lower-bound with a value of $\mu= 10^{-9}$ is chosen for practical considerations in numerical computations. For any value smaller than $\mu=10^{-9}$, issues in rounding errors due to machine precision might become evident. This results from computing the environment's machine epsilon and floating point precision.
\begin{figure*}[t]
	\centering
	\vspace{-0.1cm}
	\hspace{0.55cm}
	\subfloat{\includegraphics[keepaspectratio=true,scale=0.5]{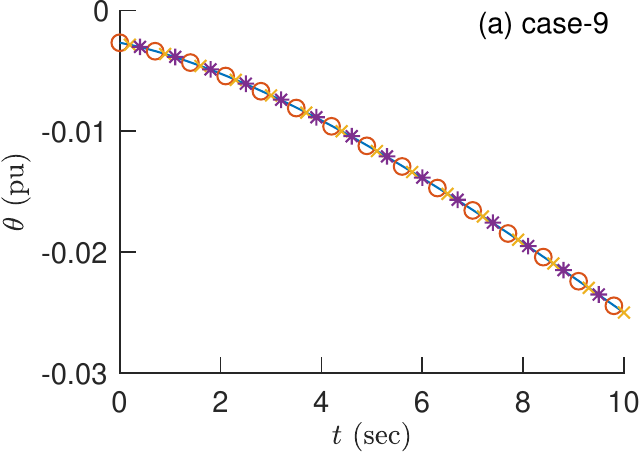}}
	\hspace{0.45cm}
	\subfloat{\includegraphics[keepaspectratio=true,scale=0.5]{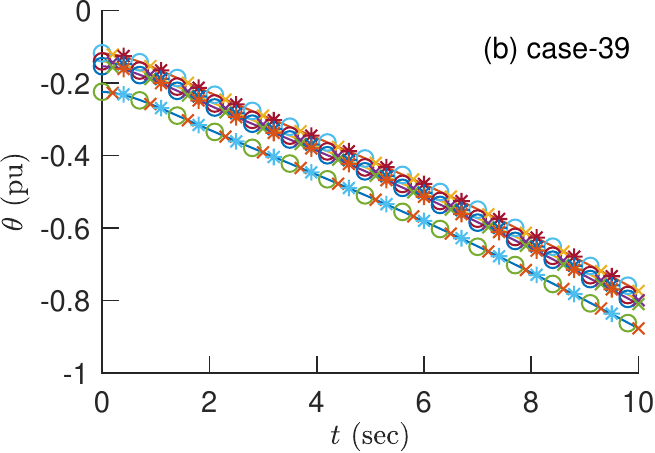}}
	\hspace{0.45cm}
	\subfloat{\includegraphics[keepaspectratio=true,scale=0.5]{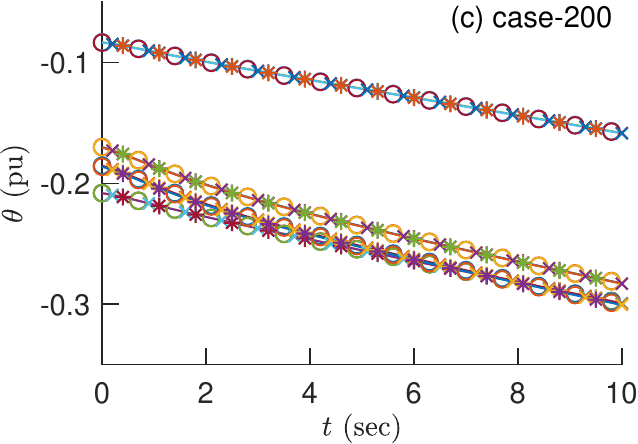}}
	\vspace{-0.6cm}
	\subfloat{\includegraphics[keepaspectratio=true,scale=0.5]{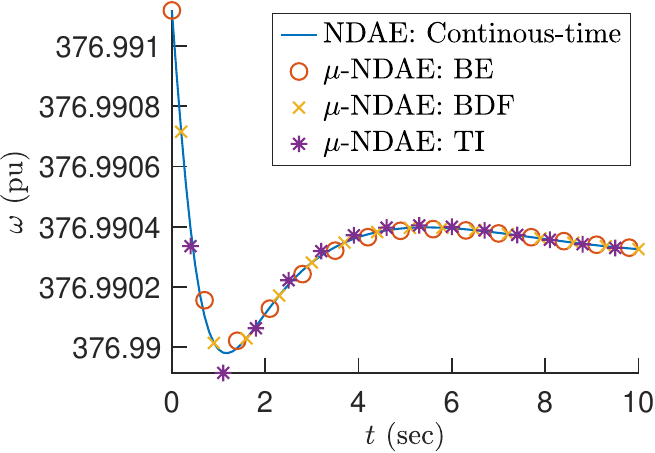}}
	\hspace{0.4cm}
	\subfloat{\includegraphics[keepaspectratio=true,scale=0.5]{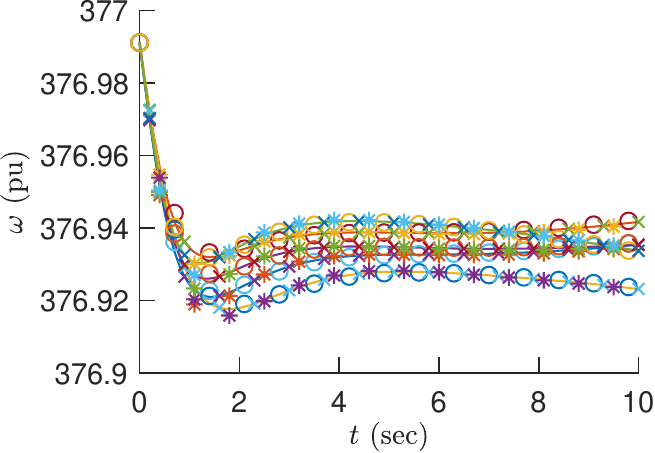}}
	\hspace{0.4cm}
	\subfloat{\includegraphics[keepaspectratio=true,scale=0.5]{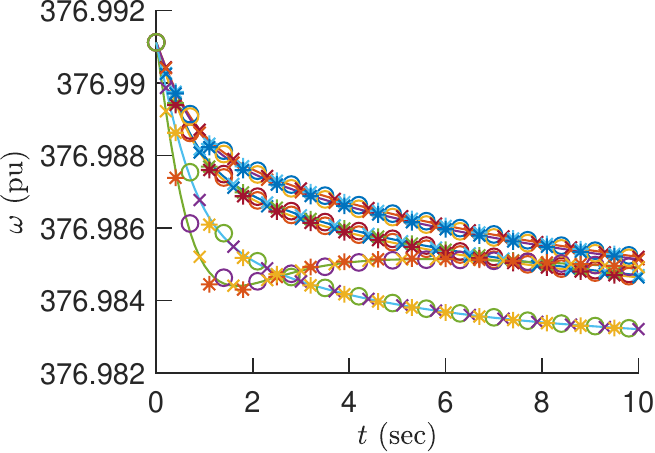}}
	\vspace{-0.1cm}
	\caption{Transient differential $(\omega_{i})$ and algebraic $(\theta_{i})$ state trajectories under load and renewables disturbance: (a) $\mr{case}$-$\mr{9}$ ($\alpha_{L} = 2\%$), (b) $\mr{case}$-$\mr{39}$ ($\alpha_{L} = 5\%$), and (c) $\mr{case}$-$\mr{200}$ ($\alpha_{L} = 20\%$). The figures represent the power system's response to load/renewables perturbations; the system returns to steady-states conditions after a certain time-span.}\label{fig:sim}
	\vspace{-0.3cm}
\end{figure*}

The $\mr{RMSE}$ of the $\mu$-NDAE model over time period $t$ is computed by considering $\m{{x}}_{k}$ corresponding to the NDAE system and $\m{\tilde{x}}_{k}$ to the $\mu$-NDAE system. The $\mr{RMSE}$ for the different cases and for each of the discretization methods over the range of $\mu$ values is depicted in Fig.~\ref{fig:disc-mu}. It can be seen that the $\mu$-NDAE system approximates the NDAE with a relatively small error and this error tends to decrease as $\mu$ approaches zero, which is intuitive since with $\mu$ reaching zero, the system reverts back to a NDAE representation. From Fig.~\ref{fig:disc-mu} one can discern that the $\mr{RMSE}$ becomes less than $10^{-3}$ when $\mu$ reaches $10^{-6}$ for the different discretization methods, in particular, for that of TI it exhibits an asymptotic behavior when $\mu \geq 10^{-6}$. The accuracy of the proposed $\mu$-NDAE model has a clear relation with the value of $\mu$ chosen. The trends for each discretization method as depicted in Fig.~\ref{fig:disc-mu} show that the choice of discretization method affects the accuracy of the simulation. The mathematical relation between $\mu$ and the RMSE also depends on the NR accuracy; it can therefore be controlled from within the NR iteration tolerance. We note that there is an approximately linear relation between the value of $\mu$ and the RMSE. The mathematical proof of such relation is outside the scope of this paper.

Having provided experimental validation of the accuracy and stability of the proposed discrete-time $\mu$-NDAE model, for the remainder of this work we choose $\mu=10^{-6}$ as it produces a sufficient approximation of the NDAE model. The differential and algebraic transient state trajectories of the studied power system cases under the different discretization methods and for $\mu=10^{-6}$ are presented in Fig.~\ref{fig:sim}. The system is simulated under transient conditions as outlined in Section~\ref{sec:OPP_Frame} (see, Step $1-3$) and the chosen load/renewables disturbance for each case system is shown in Fig.~\ref{fig:sim}. The trajectories show an accurate depiction of state-trajectories under the different discretization methods as compared with the baseline NDAE model.

\subsection{Load/Renewables Uncertainty and Measurement Noise}\label{sec:CASE-PPP}
We now solve the OPP problem, posed as~\textbf{P5}, according to the framework outlined in Section~\ref{sec:OPP_Frame}.
We solve the OPP problem for each of the test cases and under the different discretization methods while being constrained by the number of PMUs $p$ that is to be employed within the network. The maximum number of PMUs to be installed for each of the test cases is taken as $p = \mr{ceil}(\eta \times n_p)$, where PMU bus ratio $\eta \in \{0.2, 0.4, 0.6, 0.8,1\} $. That is, for example while considering $\mr{case}$-$\mr{9}$ the number of potential PMU buses is $n_p =9$. Thus, the required number of PMUs to place within the network is $p=2$ for a ratio $\eta =0.2$. A PMU ratio $\eta = 1$ means that PMUs are to placed on all potential buses within the network, that is, $p=n_p$. The OPP considering both generator and load buses locations for each of the test cases and under different discretization methods are given in Fig.~\ref{fig:case_200-OPP} (Appendix~\ref{apdx:OPP}).
 
Three key aspects can be pointed out from the proposed observability-based OPP problem. The first is that through the coupling of dynamics and algebraic states, load buses are selected and thus are included in the optimal set $\mathcal{{\m{Z}^{*}}}$, meaning that $\mathcal{{\m{Z}^{*}}} \subseteq \mathcal{N}_p$. This aspect is important since typically, when considering an ODE system representation, only generator buses are considered as potential locations for the observability-based OPP problem. This validates the rationale behind utilizing the complete NDAE representation, instead of an ODE model, of a power system for solving the observability-based OPP problem.
The second is that for each of the discretization methods that the OPP is built upon, a different optimal set is obtained. This can be clearly identified from Fig.~\ref{fig:case_200-OPP} for $\mr{case}$-$\mr{9}$. The reason for such behavior is that each of the discretization methods change the structure of the observability matrix~\eqref{eq:obs_gram}. This is a result of the different partial derivatives structures that have been derived and are presented for each of the discretization methods in Appendices~\ref{apdx:calc_Jac} and~\ref{apdx:Jac_Gear}. One can notice that the TI method differs from both BDF and BE by having an additional evaluation of the partial derivative of the system nonlinearities for an additional previous time-step. Whereas BE and BDF differ by having $k_g$ order of previous time-step dependency. We point out that the placements for BE and BDF are less different from those as compared to TI method. This similarity between BE and BDF optimal placements can be observed for $\mr{case}$-$\mr{200}$. The results suggest that indeed, the discretization method does affect the optimal PMU placements. We note here, that nonlinear models of observability depend on the operating point and the simulation model. That is, the nonlinear observability Gramian depends on the underlying structure of nonlinearities that are depicted in a dissimilar manner amongst the different discretizations.
In order to understand and assess the resulting optimal placements for each of the methods, we compare the resulting estimation error on both differential and algebraic states in Section~\ref{sec:Robustness}.

The third aspect is that it is evident that modularity is retained with the increase of PMUs selected. As depicted in Fig.~\ref{fig:case_200-OPP}, there exists a continuity of the horizontal histograms---that shows the location of the bus being selected---as the number of PMU placement increases. This means that as we increase the number of PMUs $p$ required to be employed within the network, the previous optimal set $\mathcal{{\m{Z}^{*}}}$ becomes a subset of the new OPP. This concurs with the modularity concept that the \textit{a priori} optimization problem \textbf{P5} is based upon. The significance of having a modular PMU or sensor placement framework is two-folds. First, $(i)$ with increased penetration of fuel-free energy sources---wind plants and solar farms---achieving an observable system simply requires the same grid phasor measurements or an additional PMU that augments the preexisting PMUs. This enables expanding grid operations while retaining system observability and control. Second, $(ii)$ the scheduling of PMUs or sensors can be easily performed by activating an incremental set of sensors $\Delta\mathcal{{\m{Z}^{*}}}$. This offers a fast selection approach when placing sensors or PMUs for DSE under additional physical constraints, such as cost or availability of sensors. The applicability of expanding the power network to include detailed models of renewables is to be explored in future work while considering the proposed OPP framework. On such note, we now assess the robustness of the optimal PMU placements against load/renewable perturbations and measurement noise.

\begin{figure}[t]
	\centering
	\includegraphics[keepaspectratio=true,scale=1]{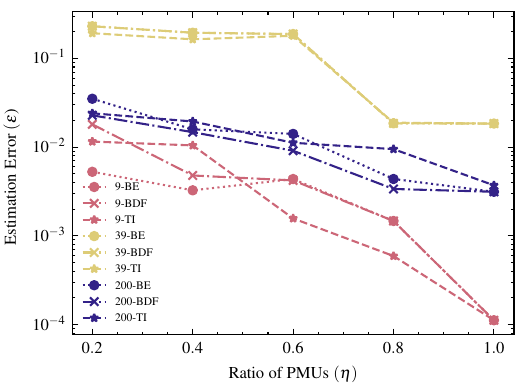}
	\vspace{-0.2cm}
	\caption{Estimation error $\eps$ resulting from the optimal PMU placements for each case and discretization methods.}\label{fig:est_error}
	\vspace{-0.25cm}
\end{figure}

\subsubsection{Effect of Measurement Noise}\label{sec:case_noise}
To investigate the impact of Gaussian measurement noise $v$ on the optimal PMU placements, we apply a noise with mean zero and $v$ in~\eqref{disc_m_NDAE} within the range of $v = \{0 \%, 5\%\}$. Then, we perform the OPP framework for each of the PMU ratios $\eta$. The resulting optimal placements $\mathcal{{\m{Z}^{*}}}$ show robustness towards measurement noise, meaning that for each case system, the same optimal placements are obtained. As such, under the studied range of noise $v$, the optimal placements $\mathcal{{\m{Z}^{*}}}$ remain the same as in Fig.~\ref{fig:case_200-OPP} for each of the cases. This robustness can be explained by how the MHE estimation framework accounts for noisy measurements, such that the observability matrix is based on estimated measurements under noise from the MHE algorithm.

\subsubsection{Effect of Load/Renewables Uncertainty}\label{sec:case_uncertainty}
As for investigating the impact of load/renewables uncertainty on the optimal placements, we vary $\alpha_{L}$ for $\mr{case}$-$\mr{9}$ and $\mr{case}$-$\mr{39}$ within the range of  $\alpha_{L}= \{0 \%, 5\%\}$ and for $\mr{case}$-$\mr{200}$ within the range of $\alpha_{L}= \{0 \%, 20\%\}$ (see, Section~\ref{sec:OPP_Frame}). The results show consistent placements when varying the perturbation values on the load and renewables, i.e., same placements as in Fig.~\ref{fig:case_200-OPP}. This is also explained by the framework that the OPP is based upon, such that the major assumption is that initial states and load/renewables disturbances are not known. Thus, we start by assuming such conditions. Then, using the MHE Gauss-Newton algorithm the actual states under the actual loads/renewables are estimated. Then, based on these estimated states the observability matrix is constructed. This means that inherent within the construction of the observability matrix, such uncertainty is already accounted for and therefore offers robustness towards load/renewables perturbations.

\subsection{Initial State Estimation under Optimal PMU Placement}\label{sec:Robustness}
We perform initial state estimation based on the optimal PMU configuration chosen for each of the cases. Note here that the estimation error $\eps$ evaluates the OPP, where a smaller $\eps$ suggests that the given PMU placements provide a more accurate reconstruction of the initial state. The estimation error for each case and under each of the discretization methods is presented in Fig.~\ref{fig:est_error}. Intuitively, as we increase the number of PMUs placed, the estimation error decreases. This relates to the concept of observability where, the more PMUs are employed or more nodes are being sensed, reconstruction of the initial dynamics and algebraic states becomes more accurate. This can be seen for each of the network cases in Fig.~\ref{fig:est_error} with the increase in ratio of PMUs being selected. 

For the case of PMUs placed, that is between $\{0.2,0.4\}$, BDF discretization estimation outperforms those of BE and TI. However, the TI method results in better estimates under increased PMU fractions. Herein, since we want to limit the number of PMUs, we consider the fractions that are small, i.e., employ lower PMUs whilst achieving adequate DSE. 
Thus based on the concept of observability and initial state reconstruction, we refer to BDF discretization method to solve for optimal PMU placements.

\subsection{Comparative Analysis: Observability-based OPP}\label{sec:Comp_Analysis}
\begin{figure}[t]
	\centering	\includegraphics[keepaspectratio=true,scale=0.7]{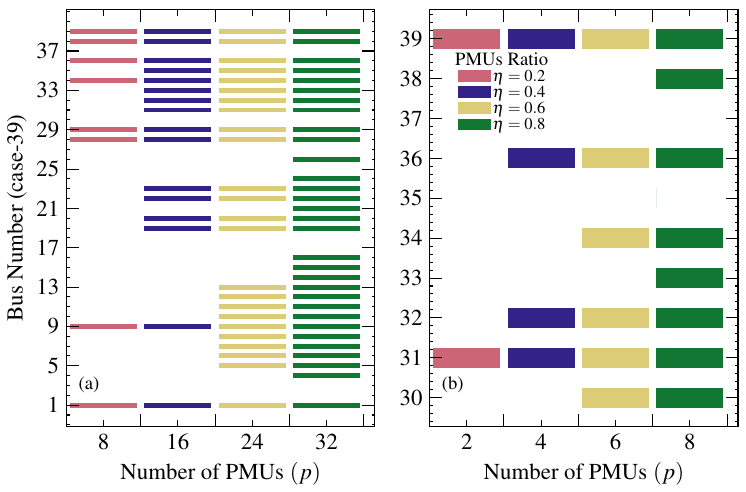}
	\caption{Comparing optimal PMU placements resulting from $(a)$ the proposed OPP problem \textbf{P5} and $(b)$ the OPP problem developed in~\cite{Qi2015}.}\label{fig:Comp_Plot}
	\vspace{-0.25cm}
\end{figure}
We now briefly compare the OPP framework herein with the observability-based OPP literature discussed in Section~\ref{sec:Introduction}. Note that in our work, we are concerned with solving the OPP from a systems theory perspective. Specifically, we approach proposing an OPP framework that is based on quantitative measures of system observability, i.e., $\mathrm{trace}(\m W(\cdot))$. The typical methods for quantifying observability as discussed in Section~\ref{sec:Introduction} include: studying the observability of the linearized system about an operating point~\cite{Qi2015,Rouhani2017}, computing the nonlinear empirical Gramian that is based on impulse response~\cite{Qi2015,Qi2016}, and building an observability matrix from Lie derivatives computations~\cite{Rouhani2017}. In~\cite{Qi2015,Rouhani2017}, both empirical Gramian and Lie derivative-based observability quantification have been compared to a linearized approach. In both studies, it is shown that nonlinear observability quantification is indeed more accurate and robust than the linearized method. Moreover, although accurate, solving an OPP problem that is based on Lie derivatives is computationally challenging and impractical. Thus, herein we compare our proposed OPP approach to the empirical Gramian OPP framework developed in~\cite{Qi2015}.

With that in mind, based on the OPP approach from~\cite{Qi2015}, we consider the $4^{th}$-order model~\eqref{eq:differential_dynamics}, while neglecting the algebraic constraints of the system. As such, the empirical Gramian is constructed based on steady-states initial conditions, since as compared to our approach, there is no clear way to directly model the load/renewables active power injections into the system. The discrete-time empirical observability matrix ${\m{W}}_{o}^{\delta}(\m{x}_{0})\in \mathbb{R}^{n \times n}$ can be written as~\eqref{eq:Emp0bsGram}, refer to~\cite{Krener2009,Qi2015,Qi2016} for more details.

\begin{equation}\label{eq:Emp0bsGram}
{\m{W}}_{o}^{\delta}(\m{x}_{0}) := \frac{1}{4\delta^{2}}\sum_{k=0}^{\mr{N}-1}(\m{\Delta Y}^{\delta}_{k})^{\top}
	\m{\Delta Y}^{\delta}_{k},
\end{equation}
where the impulse response measurement vector $\m{\Delta Y}^{\delta}_{k}:=\m{\Delta Y}^{\delta}_{k}(\m{x}_{0}) = \bmat{\m{y}_k^{+1} - \m{y}_k^{-1},\; \dots \; , \m{y}_k^{+n} - \m{y}_k^{-n}}^{\top} \in \mathbb{R}^{n_p\times n}$ and
$\m{y}_k^{\pm i} = \m{y}_k(\m{x}_{0}\pm \delta\m{e}_{i},i) \in \mathbb{R}^{1 \times n_p}$. Here $\m{x}_o$ considers only the dynamic states. The Gramian~\eqref{eq:Emp0bsGram} is based on initial state impulse response, where $\m{e}_i \in \mathbb{R}^{n} $ for $i=1,\dots,n$ denotes the standard basis vector and $\delta> 0$ is a constant positive infinitesimal parameter. The empirical observability matrix is computationally expensive; it requires simulating the power system from $2n$ impulse response initial conditions over the observation horizon $\mr{N}_o$. We note here, that the measurements $\m{y}_k$ consider measuring only $\{\m{v}, \m{\theta}\}$. The OPP problem is then posed as \textbf{P3}; it is considered as a nonconvex mixed-integer nonlinear program. In~\cite{Qi2015}, the NOMAD solver is used to solve \textbf{P3}, however, NOMAD can not be interfaced with MATLAB running on an Apple M1 pro chip. As such, we solve \textbf{P3} using the BNB algorithm with Gurobi as the solver. For additional information regarding the empirical Gramian OPP problem refer to~\cite{Qi2015}.

The optimal PMU placements resulting from the OPP approach in~\cite{Qi2015} in comparison to \textbf{ P5} for $\mr{case}$-$\mr{39}$ are depicted in~\ref{fig:Comp_Plot}. Three fundamental aspects of our approach as compared to the empirical Gramian approach, that only considers the differential system model~\eqref{eq:differential_dynamics}, can be identified. First, by considering the algebraic constraint, i.e., NDAE system model, the potential buses for the PMU selection includes load buses, whereas that of~\cite{Qi2015} considers only generator buses numbered $30 \rightarrow 39$. Second, load/renewables uncertainty is not considered. As mentioned in~\cite{Qi2015}, robustness against transient loads is checked after obtaining the optimal PMU locations. Third, PMU placement modularity under the empirical Gramian approach is not retained. Check generator bus numbers $32$, $34$, and $39$ for PMU ratios $\eta = \{0.6, 0.8\}$ in Fig.~\ref{fig:Comp_Plot}. This means that adding additional PMUs to an existing optimal set of PMUs might require the removal of PMUs from buses already installed. This is known as multi-stage PMU placement and is an important aspect to study in the context of power networks that are ever-growing.

\section{Paper summary, Limitations and Future Work}\label{sec:summary}
This paper revisits the OPP problem for a network of power systems. The power system is based on a NDAE representation which allows coupling of the differential and algebraic states within the network. The NDAE system is discretized using BE, BDF, and TI discretization methods and is transformed into a $\mu$-NDAE which retains the mathematical structure of an ODE. We adopt a MHE approach to perform the OPP problem by exploiting the modularity of the observability matrix. As such, we pose the OPP as an~\textit{a priori set optimization program} which extenuates the computational burden of performing complex computation at each optimization instance when solving the combinatorial placement problem. Given the above comprehensive investigation and validation, we answer the posed research questions in Section~\ref{sec:case_study}:
\begin{itemize}[leftmargin=*]
	\item \textbf{A1}: There is a value for $\mu$ the ensures numerical stability and solvability of the NDAE power system. Such value bounds the error on differential and algebraic state trajectories.
	\item \textbf{A2}: The choice of discretization method is important when formulating the nonlinear observability-based approach. This depends on the system nonlinearities and their structure. For the power system herein, BDF method is suggested for the OPP problem.
	\item \textbf{A3}: Indeed, robustness against measurement noise and load/renewables uncertainty is achieved under the OPP framework. This is inherent with the MHE framework that the observability-based OPP is built upon.
\item \textbf{A4}: Modularity of the optimal PMU placements is observed when increasing the number of PMUs to be employed within the network. As mentioned in Section~\ref{sec:CASE-PPP}, this offers computationally efficient solutions for ever-growing power grids.
\end{itemize}
The paper is limited to modeling load/renewables as constant power injections. A more detailed representation of power-electronic systems, such as solar and wind, is essential. Accordingly, our future work will include understanding how system observability and OPP are affected with the allocation of new renewable resources to the power network. In specific, we want to understand if modular multi-stage PMU placement is still applicable when modeling power-electronics along with their uncertainty.

\bibliographystyle{IEEEtran}
\bibliography{OSP_for_NDAE}

\appendices
\section{Paper Nomenclature}\label{apdx:nom}
Tab.~\ref{tab:notation} represents the paper nomenclature.

\section{Implicit Discrete-Time Modeling}\label{apdx:Imp_Disc}
In this section, we present the BE and TI discrete-time models.
We define the discretization constants $\tilde{h}$ equal to $h$ for BE, $\beta h$ for BDF, and $0.5 h$ for TI method.
	
\subsubsection{Backward Euler} Since the BDF method is a generalization of BE's discretization method, we can take the index $k_g = 1$. Then, the $\mu$-NDAE system can be discretized as

\begin{subequations}\label{eq:backward_euler}
	\begin{align}
		\m{x}_{d,k} - \m{x}_{d,k- 1} &=  \tilde{h}\big(\m{f}(\m{z}_{k})\big), \\
		\mu\m{x}_{a,k} - \mu \m{x}_{a,k- 1}& = \tilde{h}\big(\m{g}(\m{x}_{k})\big),
	\end{align} 
\end{subequations}
and the system dynamics under BE method of the $\mu$-NDAE system in a state-space representation can be written as 
\begin{equation}\label{eq:mu-SS-BE}
	{\mE_{\mu}}{\m{x}_{k}} = {\mE_{\mu}}{\m{x}_{k-1}} + \tilde{h}\eye_{n}
	\begin{bmatrix}
		\m{f}(\m{z}_{k})
		\\ \m{g}(\m{x}_{k})
	\end{bmatrix},
\end{equation}

\begin{table}[t]
		\renewcommand{\arraystretch}{1.42}
		\caption{Paper nomenclature: parameter, variable, and set definitions.}
		\label{tab:notation}		\vspace{-0.05cm}		
		\centering
		\vspace{-0.3cm}
		\begin{tabular}{|l|p{\dimexpr\columnwidth-4\tabcolsep-3\arrayrulewidth-1.5cm}|}
			\hline
			\textbf{Notation} & \textbf{Description}		\vspace{-0.2cm}		\\ 
			\hline
			\hline
			\hspace{-0.1cm}$\mathcal{N}$ & \hspace{-0.1cm}set of nodes (buses)\\
			\hline
			\hspace{-0.1cm}$\mathcal{E}$ & \hspace{-0.1cm}set of transmission lines\\
			\hline
			\hspace{-0.1cm}$\mathcal{G}$ & \hspace{-0.1cm}set of generator buses, $\mathcal{G} = \{1,2, \dots,G\}$\\
			\hline
			\hspace{-0.1cm}$\mathcal{L}$ & \hspace{-0.1cm}set of load buses, $\mathcal{L} = \{1,2, \dots,L\}$\\
			\hline
			\hspace{-0.1cm}$\mathcal{R}$ & \hspace{-0.1cm}set of renewable buses, $\mathcal{R} = \{1,2, \dots,R\}$\\
			\hline
			\hspace{-0.1cm}$\delta_{i}$ & \hspace{-0.1cm}generator rotor angle $\mr{(rad)}$\\
			\hline
			\hspace{-0.1cm}$\omega_0$ & \hspace{-0.1cm}generator synchronous speed $(120\pi \;  \mr{rad/sec})$\\
			\hline
			\hspace{-0.1cm}$\omega_i$ & \hspace{-0.1cm}generator rotor speed$\mr{(rad/sec)}$\\
			\hline
			\hspace{-0.1cm}${E}^{'}_{i}$  & \hspace{-0.1cm}generator transient voltage $\mr{(pu)}$\\
			\hline
			\hspace{-0.1cm}${T}_{\mr{M}i}$ & \hspace{-0.1cm}generator mechanical torque $\mr{(pu)}$\\
			\hline
			\hspace{-0.1cm}$E_{\mr{fd}i}$ & \hspace{-0.1cm}generator internal field voltage $\mr{(pu)}$\\
			\hline
			\hspace{-0.1cm}$T_{\mr{r}i} $ & \hspace{-0.1cm}governor reference signal $\mr{(pu)}$\\
			\hline
			\hspace{-0.1cm}$M_{i}$ & \hspace{-0.1cm}rotor inertia constant $(\mr{pu} \times \mr{sec}^2)$\\
			\hline
			\hspace{-0.1cm}$D_i$ & \hspace{-0.1cm}damping coefficient $(\mr{pu} \times \mr{sec}^2)$\\
			\hline
			\hspace{-0.1cm}$x_{\mr{d}i}$ and $x_{\mr{q}i}$ & \hspace{-0.1cm}synchronous reactance $\mr{(pu)}$\\
			\hline
			\hspace{-0.1cm}$x^{'}_{\mr{d}i}$ & \hspace{-0.1cm}direct-axis transient reactance $\mr{(pu)}$\\
			\hline
			\hspace{-0.1cm}$T^{'}_{\mr{d0}i}$ & \hspace{-0.1cm}direct-axis open-circuit time constant $\mr{(sec)}$\\
			\hline
			\hspace{-0.1cm}$T_{\mr{CH}i}$ & \hspace{-0.1cm}chest valve time constant $\mr{(sec)}$\\
			\hline
			\hspace{-0.1cm} $R_{\mr{D}i}$ & \hspace{-0.1cm}governor regulation constant $\mr{(Hz/pu)}$\\
			\hline
			\hspace{-0.1cm}$P_{\mr{G}i}$ and $Q_{\mr{G}i}$ & \hspace{-0.1cm}generator's real and reactive power  $\mr{(pu)}$\\
			\hline
			\hspace{-0.1cm}$P_{\mr{R}i}$ and $Q_{\mr{R}i}$ & \hspace{-0.1cm}renewable's active and reactive power $\mr{(pu)}$\\
			\hline 		
			\hspace{-0.1cm}$P_{\mr{L}i}$ and $Q_{\mr{L}i}$ & \hspace{-0.1cm}load's active and reactive power $\mr{(pu)}$\\
			\hline 		
			\hspace{-0.1cm}$\theta_{ij}$ & \hspace{-0.1cm}bus angle, $\theta_{ij} = \theta_{i} - \theta_{j}$ $\mr{(pu)}$\\
			\hline
			\hspace{-0.1cm}$v_i$ & \hspace{-0.1cm}bus voltage $\mr{(pu)}$\\
			\hline
			\hspace{-0.1cm}$G_{ij}$ and $B_{ij}$ & \hspace{-0.1cm}conductance and susceptance between bus $i$ and $j$ \\
			\hline
		\end{tabular}
	\end{table}
	
\subsubsection{Trapezoidal Implicit} For the TI discretization method, the $\mu$-NDAE system can be written as
\begin{subequations}\label{eq:mu-TI_discre}
	\begin{align}
		\m{x}_{d,k} - \m{x}_{d,k-1} &=  \tilde{h}\big(\m{f}(\m{z}_{k}) + \m f(\m{z}_{k-1})\big),\\
	\mu	\m{x}_{a,k} - \mu \m{x}_{a,k-1} & =\tilde{h}\big( \m{g}(\m{x}_{k})+\m{g}(\m{x}_{k-1})\big),
	\end{align} 
\end{subequations}
and the system dynamics under TI method of the $\mu$-NDAE system in a state-space representation can be written as 
\begin{equation}\label{eq:mu-SS-TI}
	{\mE_{\mu}}{\m{x}_{k}} = {\mE_{\mu}}{\m{x}_{k-1}} + \tilde{h}\eye_{n}
	\begin{bmatrix}
		\m{f}(\m{z}_{k})
		+ \m{f}(\m{z}_{k-1}) \\ \m{g}(\m{x}_{k})+ \m{g}(\m{x}_{k-1})
	\end{bmatrix},
\end{equation}

Following such discretization schemes, we move forward with solving the system using NR algorithm analogously to that of Gear's method with the exception of having different Jacobian matrices under the iteration increment update.

The Jacobian $\m{A}_{g}(\cdot)$ of the implicit BE nonlinear system has the same representation as of that of Gear's method. On the other hand, the Jacobian $\m{A}_{g}(\cdot)$ of the implicit TI nonlinear system can be written as 
\begin{equation}\label{eq:mu-TI-Jac_Newton_Raph}
	\hspace{-0.3cm}	\m{A}_{g}(\m{z}^{(i)}_{k})=
	\begin{bmatrix}
		\eye_{n_d}-\tilde{h} \tilde{\m{F}}_{\m{x}_{d}}(\m{z}^{(i)}) & -\tilde{h} \tilde{\m{F}}_{\m{x}_{a}}(\m{z}^{(i)}) \\
		-\tilde{h}\tilde{\m{G}}_{\m{x}_{d}}(\m{x}^{(i)}) & \mu \eye_{n_a} -\tilde{h} \tilde{\m{G}}_{\m{x}_{a}}(\m{x}^{(i)})
	\end{bmatrix},
\end{equation}
where matrix $\tilde{\m{F}}_{\m{x}_d} (\m z^{(i)}):=  \m{F}_{\m{x}_d} (\m z^{(i)}_{k})+\m{F}_{\m{x}_d} (\m z^{(i)}_{k-1})$, matrix $\tilde{\m{F}}_{\m{x}_a} (\m z^{(i)}):=  \m{F}_{\m{x}_a} (\m z^{(i)}_{k})+\m{F}_{\m{x}_a} (\m z^{(i)}_{k-1})$, matrix $\tilde{\m{G}}_{\m{x}_{d}}(\m x^{(i)}) := \m{G}_{\m{x}_{d}}(\m x^{(i)}_{k})+\m{G}_{\m{x}_{d}}(\m x^{(i)}_{k-1})$, 
and matrix $\tilde{\m{G}}_{\m{x}_{a}}(\m x^{(i)}) := \m{G}_{\m{x}_{a}}(\m x^{(i)}_{k})+\m{G}_{\m{x}_{a}}(\m x^{(i)}_{k-1})$.
	
\section{Simulating the NDAE Discrete-time Power System}\label{apdx:NDAE-disc}
In this section, we formulate the methodology for solving the discrete-time NDAE system that is used as a basis to validate the $\mu$-NDAE system under which the OPP problem is posed.

Applying Gear's method to the discrete-time NDAE system~\eqref{eq:semi_NDAE_rep}, the dynamics of the system represented in state-space form can be rewritten  as

\begin{equation}\label{eq:discrete_NDEA_concat}
	\mE \m{x}_{k} = \mE \sum_{s=1}^{k_g}\alpha_s \m{x}_{d,k-s}
	+ \begin{bmatrix}
		\tilde{h} && 0 \\
		0 && 1
	\end{bmatrix} \begin{bmatrix}
		\m{f}(\m{z}_{k})\\
		\m{g}(\m{x}_{k})\\
	\end{bmatrix},
\end{equation}
where $\m{E} \in \mathbb{R}^{n\times n}$ is a singular diagonal matrix of $\mr{rank} = n_d$, such that, $\mE(i,i) = 1$ for $1 \leq i \leq n_{d}$ and $\mE(i,i) = 1$ for $n_{d+1} \leq i \leq n$. We solve the BDF system using NR algorithm and represent the discrete-time model under NR iteration index $i$ as follows

\begin{equation}\label{eq:discrete_NDEA_Implicit}
	\m{\phi}(\m{z}^{(i)}_{k})= \m{x}^{(i)}_{d,k} -\sum_{s=1}^{k_g}\alpha_s \m {x}_{d,k-s} - \tilde{h}\m{f}(\m{z}^{(i)}_{k}),
\end{equation}
where the BDF discrete-time NDAE system \eqref{eq:discrete_NDEA_concat} can be implicitly and succinctly written as

\begin{equation}\label{eq:discrete_NDEA_Implicit_full}
		\m{0}=  \m{\phi}(\m z^{(i)}_{k}), \quad 
		\m{0}= \m{g}(\m x^{(i)}_{k}).
\end{equation}

Following the same methodology for simulating $\mu$-NDAE system, the Jacobian $A_{g}(\cdot)$ of the implicit NDAE system~\eqref{eq:discrete_NDEA_Implicit_full} can be written as

\begin{equation}\label{eq:Jac_Newton_Raph}
	\begin{split}
		\hspace{-0.3cm}\m{A}_{g}(\cdot) &\hspace{-0.05cm}=\hspace{-0.05cm} \begin{bmatrix}
			\frac{\partial \m{\phi}(\m z^{(i)}_{k})}{\partial \m x_d} \hspace{-0.05cm}&\hspace{-0.05cm} \frac{\partial \m{\phi}(\m z^{(i)}_{k})}{\partial \m x_a} \\ 
			\frac{\partial \m{g}(\m x^{(i)}_k)}{\partial \m x_d} \hspace{-0.05cm}&\hspace{-0.05cm} \frac{\partial \m{g}(\m x^{(i)}_{k})}{\partial \m x_a} \\
		\end{bmatrix}\\
	&	\hspace{-0.05cm} =\hspace{-0.05cm} \begin{bmatrix}
			\eye_{n_d} - \tilde{h} \m{F}_{x_{d}}(\m z^{(i)}_{k}) \hspace{-0.05cm}& \hspace{-0.05cm}-\tilde{h} \m{F}_{x_{a}}(\m z^{(i)}_{k})\\
			\m{G}_{x_{d}}(\m x^{(i)}_{k}) \hspace{-0.05cm}&\hspace{-0.05cm} \m{G}_{x_{a}}(\m x^{(i)}_{k})
		\end{bmatrix},\hspace{-0.2cm}
		\end{split}
\end{equation}
then for each iteration $i$ the NR increment is computed from~\eqref{eq:newton_raph}. This increment is then used to update $\m x^{(i+1)}$ in~\eqref{eq:new_newton_raph} until convergence is satisfied. Convergence is calculated using the $\mathcal{L}$-$2$--norm on the increment.
\begin{subequations}\label{eq:newton_raph}
	\begin{align}
		\Delta \m x^{(i)}_{k} = [\m A_{g}(\m z^{(i)}_{k})]^{-1}\begin{bmatrix}
			\m{\phi}(\m z^{(i)}_{k}) \\
			\m{g}(\m x^{(i)}_{k})
		\end{bmatrix},
	\end{align} 
\end{subequations}
\begin{subequations}\label{eq:new_newton_raph}
	\begin{align}
		\m x^{(i+1)}_{d,k} = \m x^{(i)}_{d,k} + \Delta \m x^{(i)}_{d,k},\\
		\m x^{(i+1)}_{a,k} = \m x^{(i)}_{a,k} + \Delta \m x^{(i)}_{a,k} .
	\end{align} 
\end{subequations}
For brevity, we only present the NR method for BDF discretization. 

{\section{MHE Algorithm}}\label{apdx:alg_MHE}
{The following algorithm details the moving horizon estimation framework discussed in Section~\ref{sec:Initial-State-Estimation}.}

\begin{algorithm}[h]
	\caption{MHE via Gauss-Newton Iterations}\label{alg:Disc-Alg}
	\KwIn{$h_g$, $\m{x}_{0}$, $\m{u}_{0}$, $\m{\Gamma}$, tolerance}
	\KwOut{$\hat{\m{x}}_0$}
	Set $i = 1$ as GN iteration index\\
	\While {$\mathcal{L}_2$--norm of the residual $\;\geq\;$ tolerance}{Simulate the system dynamics with initial states $\m{x}_{0}$\\Build the residual function $\m{r}(\m{q})$ represented in~\eqref{eq:residual}\\Calculate the Jacobian $\m{J}_g(\m{q}^{(i)})$ in~\eqref{eq:Jac_Gauss}\\Perform the GN iteration update on $\m{q}^{i+1}$ in~\eqref{eq:Gauss-itr}\\ Update GN iteration index $i=i+1$\\ Update initial states $\m{x}_{0} \rightarrow \hat{\m{x}}_{0}$\\Calculate $\mathcal{L}_2$--norm of the residual \eqref{eq:residual}
	}	
\end{algorithm}

\section{Explicit Representation of the Jacobian Matrix}\label{apdx:calc_Jac}
We present the methodology that allows us to compute the Jacobian matrix of the power system. The chain rule is then utilized to express the Jacobian matrix~\eqref{eq:Jac_obs_horz} for an observation horizon $\mr{N}_o$. Given the implicit discrete-time models used to simulate the system, we run into a challenge of explicitly representing $\frac{\partial \m{x}_{j}}{\partial \m{x}_{j-1}}$. We herein show the derivations and calculations required to obtain an explicit representation of $\frac{\partial \m{x}_{j}}{\partial \m{x}_{j-1}}$ for computing $\m{J}(\cdot)$.

For the BDF method, the partial derivative can be obtained by taking the partial derivative of the discretized system~\eqref{eq:discrete_NDEA_concat} as follows
\begin{equation}\label{eq:gear_disc_partial}
	\mE \frac{\partial \m{x}_{j}}{\partial \m{x}_{j-1}} = \mE \sum_{s=1}^{k_g}\alpha_s \frac{\partial \m{x}_{j-s}}{\partial \m{x}_{j-1}} + \begin{bmatrix}
		\tilde{h} & 0 \\
		0 & 1
	\end{bmatrix} \begin{bmatrix}
		\frac{\partial \m{f}(\m{x}_{j})}{\partial \m{x}_{j}}|_{\m{x}_j}\frac{\partial \m{x}_{j}}{\partial \m{x}_{j-1}}\\
		\frac{\partial \m{g}(\m{x}_{j})}{\partial \m{x}_{j}}|_{\m{x}_j}\frac{\partial \m{x}_{j}}{\partial \m{x}_{j-1}}\\
	\end{bmatrix},
\end{equation}
which can equivalently be written as 
\begin{subequations}\label{eq:gear_disc_partial1}
	\begin{align}
		\frac{\partial \m{x}_{d,j}}{\partial \m{x}_{j-1}} &= \alpha_{s(1)}\frac{\partial \m{x}_{d,j-1}}{\partial \m{x}_{j-1}} + \tilde{h}  \frac{\partial \m{f}(\m{x}_{j})}{\partial \m{x}_{j}}|_{\m{x}_j}\frac{\partial \m{x}_{j}}{\partial \m{x}_{j-1}},\label{eq:gear_disc_partial1_1}\\ 
		\m{0} &= \m{0}+
		\frac{\partial \m{g}(\m{x}_{j})}{\partial \m{x}_{j}}|_{\m{x}_j}\frac{\partial \m{x}_{j}}{\partial \m{x}_{j-1}}.\label{eq:gear_disc_partial1_2}
	\end{align}
\end{subequations}

It is evident from~\eqref{eq:gear_disc_partial1} that $\frac{\partial \m{x}_{j}}{\partial \m{x}_{j-1}}$ for the case of differential variables in equation \eqref{eq:gear_disc_partial1_1} can be explicitly formulated assuming that $ (\eye+ \tilde{h}  \frac{\partial \m{f}(\m{x}_{j})}{\partial \m{x}_{j}}|_{\m{x}_j}\frac{\partial \m{x}_{j}}{\partial \m{x}_{j-1}})$ is invertible. However, for the algebraic constraint variables in equation \eqref{eq:gear_disc_partial1_2}, $\frac{\partial \m{x}_{j}}{\partial \m{x}_{j-1}}$  is equal to zero and cannot be explicitly represented. This is where the $\mu$-NDAE formulation comes into play to allow for a plausible solution towards developing an observability-based OPP framework for power systems of index-1.

To construct the Jacobian in~\eqref{eq:Jac_obs_horz}, we represent the partial derivative $\frac{\partial \m{x}_{j}}{\partial \m{x}_{j-1}}$ explicitly for each of the discretization methods presented in this work. The partial derivative $\frac{\partial \m{x}_{j}}{\partial \m{x}_{j-1}}$ for each of the discretization method can be written as

\begin{equation}\label{eq:delx_x}
	\frac{\partial \m{x}_{j}}{\partial \m{x}_{j-1}} =
	\begin{bmatrix}
		\frac{\partial \m{x}_{d,j}}{\partial \m{x}_{d,j-1}} & \frac{\partial \m{x}_{d,j}}{\partial \m{x}_{a,j-1}}\\ 
		\frac{\partial \m{x}_{a,j}}{\partial \m{x}_{d,j-1}} & \frac{\partial \m{x}_{a,j}}{\partial \m{x}_{a,j-1}}
	\end{bmatrix},\\
\end{equation}
where the partial derivative $\frac{\partial \m{x}_{j}}{\partial \m{x}_{j-1}}$ can be explicitly written for BDF method according to the following
\begin{equation}\label{mu-NDAE_state_power}
	\mE_{\mu}\frac{\partial \m{x}_{j}}{\partial \m{x}_{j-1}} = \sum_{s=1}^{k_g}\alpha_s \mE_{\mu}\frac{\partial \m{x}_{j-s}}{\partial \m{x}_{j-1}} + \tilde{h} \eye_{n} \begin{bmatrix}
		\frac{\partial \m{f}(\m{x}_{j})}{\partial \m{x}_{j}}|_{\m{x}_j}\frac{\partial \m{x}_{j}}{\partial \m{x}_{j-1}}\\
		\frac{\partial \m{g}(\m{x}_{j})}{\partial \m{x}_{j}}|_{\m{x}_j}\frac{\partial \m{x}_{j}}{\partial \m{x}_{j-1}}\\
	\end{bmatrix},
\end{equation}
such that by differentiating with respect to the differential and algebraic state variables separately, we obtain the following

\begin{subequations}\label{eq:mu-NDAE_partial_exp}
	\begin{align}
	\hspace{-0.3cm}\frac{\partial \m{x}_{d_{j}}}{\partial \m{x}_{j-1}} \hspace{-0.05cm}&=\hspace{-0.05cm} \begin{bmatrix}
			\alpha_{s(1)} + \tilde{h}  \frac{\partial \m{f}(\m{x}_{j})}{\partial \m{x}_{d_{j}}}|_{\m{x}_j}\frac{\partial \m{x}_{d_{j}}}{\partial \m{x}_{d_{j-1}}} &
			\tilde{h}  \frac{\partial \m{f}(\m{x}_{j})}{\partial \m{x}_{a_{j}}}|_{\m{x}_j}\frac{\partial \m{x}_{a_{j}}}{\partial \m{x}_{a_{j-1}}}\\ \end{bmatrix}^{\top}\hspace{-0.2cm},\\
	 \hspace{-0.3cm}\frac{\partial \m{x}_{a_{j}}}{\partial \m{x}_{j-1}} \hspace{-0.05cm}&=\hspace{-0.05cm} \begin{bmatrix}
			\tilde{h}  \frac{\partial \m{g}(\m{x}_{j})}{\partial \m{x}_{d_{j}}}|_{\m{x}_j}\frac{\partial \m{x}_{d_{j}}}{\partial \m{x}_{d_{j-1}}} & 
			\mu \alpha_{s(1)} + \tilde{h}  \frac{\partial \m{g}(\m{x}_{j})}{\partial \m{x}_{a_{j}}}|_{\m{x}_j}\frac{\partial \m{x}_{a_{j}}}{\partial 	\m{x}_{a_{j-1}}}\\ \end{bmatrix}^{\top}.\hspace{-0.2cm}
	\end{align}
\end{subequations}

we can rewrite~\eqref{eq:mu-NDAE_partial_exp} in implicit form as 

\begin{subequations}\label{eq:mu-NDAE_partial_Implicit2}
	\begin{align}
	&\begin{bmatrix}\tfrac{\partial \m{x}_{d_{j}}}{\partial \m{x}_{d_{j-1}}}\\  \tfrac{\partial \m{x}_{d_{j}}}{\partial \m{x}_{a_{j-1}}}\end{bmatrix} = \begin{bmatrix}
		\begin{bmatrix}
			\eye_{n_d} - \tilde{h}  \frac{\partial \m{f}(\m{x}_{j})}{\partial \m{x}_{d_{j}}}|_{\m{x}_j}
		\end{bmatrix}^{-1}\alpha_{s(1)}\\
		\tilde{h}  \frac{\partial \m{f}(\m{x}_{j})}{\partial \m{x}_{a_{j}}}|_{\m{x}_j}\frac{\partial \m{x}_{a_{j}}}{\partial \m{x}_{a_{j-1}}}\end{bmatrix},\\
		&\begin{bmatrix}
			\tfrac{\partial \m{x}_{a_{j}}}{\partial \m{x}_{d_{j-1}}} \\
			\tfrac{\partial \m{x}_{a_{j}}}{\partial \m{x}_{a{j-1}}}
		\end{bmatrix} = \begin{bmatrix}
		\tilde{h} \frac{\partial \m{g}(\m{x}_{j})}{\partial \m{x}_{d_{j}}}|_{\m{x}_j}\frac{\partial \m{x}_{d_{j}}}{\partial \m{x}_{d_{j-1}}}\mu^{-1}\\
		\begin{bmatrix}
			\mu \eye_{n_a} -\tilde{h}  \frac{\partial \m{g}(\m{x}_{j})}{\partial \m{x}_{a_{j}}}|_{\m{x}_j}
		\end{bmatrix}^{-1}\mu \alpha_{s(1)}
	\end{bmatrix},
	\end{align}
\end{subequations}
let $\mA_{n_d} = (\eye_{n_d} - \tilde{h}  \frac{\partial \m{f}(\m{x}_{j})}{\partial \m{x}_{d_{j}}}|_{\m{x}_j}) $ and $\mA_{n_a} = (\mu \eye_{n_a} - \tilde{h}  \frac{\partial \m{g}(\m{x}_{j})}{\partial \m{x}_{a_{j}}}|_{\m{x}_j})$, such that, $\mA_{n_d} $ and $\mA_{n_a}$ are invertible. To explicitly represent $\frac{\partial \m{x}_{d_{j}}}{\partial \m{x}_{a_{j-1}}}$ and $\frac{\partial \m{x}_{a_{j}}}{\partial \m{x}_{d_{j-1}}}$, we replace  $\frac{\partial \m{x}_{d_{j}}}{\partial \m{x}_{d_{j-1}}}$ and $\frac{\partial \m{x}_{a_{j}}}{\partial \m{x}_{a{j-1}}}$ by their explicit representation formulated in~\eqref{eq:mu-NDAE_partial_Implicit2} as follows
\begin{subequations}\label{eq:mu-NDAE_partial_Implicit3}
	\begin{align}
	\begin{bmatrix}\tfrac{\partial \m{x}_{d_{j}}}{\partial \m{x}_{d_{j-1}}}\\  \tfrac{\partial \m{x}_{d_{j}}}{\partial \m{x}_{a_{j-1}}}\end{bmatrix}&= \begin{bmatrix}
			\begin{bmatrix}
				\mA_{n_d}
			\end{bmatrix}^{-1}\alpha_{s(1)}\eye_{n_d}\\  \\
			\tilde{h}  \frac{\partial \m f(\m{x}_{j})}{\partial \m{x}_{a_{j}}}|_{\m{x}_j}\begin{bmatrix}
				\mA_{n_a}
			\end{bmatrix}^{-1}\mu \alpha_{s(1)}\eye_{n_a}\\ \end{bmatrix},\\ 
\begin{bmatrix}
	\tfrac{\partial \m{x}_{a_{j}}}{\partial \m{x}_{d_{j-1}}} \\
	\tfrac{\partial \m{x}_{a_{j}}}{\partial \m{x}_{a{j-1}}}
\end{bmatrix} &= \begin{bmatrix}
			\tilde{h} \frac{\partial \m g(\m{x}_{j})}{\partial \m{x}_{d_{j}}}|_{\m{x}_j}\begin{bmatrix}
				\mA_{n_d}
			\end{bmatrix}^{-1} \alpha_{s(1)}\eye_{n_d}\mu^{-1}\\  \\
			\begin{bmatrix}
				\mA_{n_a}
			\end{bmatrix}^{-1}\mu \alpha_{s(1)}\eye_{n_a}\\ \end{bmatrix}.
	\end{align}
\end{subequations}

Now for the case where $k_g= 1$, that is the case of Backward Euler, \eqref{eq:mu-NDAE_partial_Implicit3} can be written as

\begin{equation}\label{eq:BEJac_mu-NDAE_partial}
\hspace{-0.25cm}	\frac{\partial \m{x}_{j}}{\partial \m{x}_{j-1}}  \hspace{-0.05cm}=  \hspace{-0.05cm}
	\begin{bmatrix}
		\begin{bmatrix}
			\mA_{n_d}
		\end{bmatrix}^{-1}\eye_{n_d}& \hspace{-0.3cm}
		\tilde{h} \mF_{\m{x}_{a,j}}\begin{bmatrix}
			\mA_{n_a}
		\end{bmatrix}^{-1}\mu\eye_{n_a}\\ 
		\tilde{h} \mG_{\m{x}_{d,j}}\begin{bmatrix}
			\mA_{n_d}
		\end{bmatrix}^{-1}\eye_{n_d}\mu^{-1}& \hspace{-0.3cm}
		\begin{bmatrix}
			\mA_{n_a}
		\end{bmatrix}^{-1}\mu\eye_{n_a}\\
	\end{bmatrix},
\end{equation}
where matrix $\mF_{\m{x}_{d,j}} = \frac{\partial \m{f}(\m{x}_{j})}{\partial \m{x}_{d_{j}}}|_{\m{x}_j}$, matrix $\mF_{\m{x}_{a,j}}= \frac{\partial \m{f}(\m{x}_{j})}{\partial \m{x}_{a_{j}}}|_{\m{x}_j}$, matrix $\mG_{\m{x}_{d,j}} = \frac{\partial \m{g}(\m{x}_{j})}{\partial \m{x}_{d_{j}}}|_{\m{x}_j}$, and matrix  $\mG_{\m{x}_{a,j}} = \frac{\partial \m{g}(\m{x}_{j})}{\partial \m{x}_{a_{j}}}|_{\m{x}_j^{(i)}}$.

To represent the partial derivative $\frac{\partial \m{x}_{j+1}}{\partial \m{x}_{j}}$ explicitly for the TI discrete-time model of the system, we start by differentiating~\eqref{disc_ss_NDAE} with respect to both the differential and algebraic variables. We denote $\mF_{\m{x}_{j}} = \frac{\partial \m{f}(\m{x}_{j})}{\partial \m{x}_{{j}}}|_{\m{x}_j}$ and $\mG_{\m{x}_{j}} = \frac{\partial \m{g}(\m{x}_{j})}{\partial \m{x}_{{j}}}|_{\m{x}_j^{(i)}}$. 
Differentiating with respect to the differential and algebraic state variables separately~\eqref{disc_ss_NDAE} can be written as

\begin{subequations}\label{eq:TRAPmu-NDAE_partial_exp}
	\begin{align}
		\begin{bmatrix}\tfrac{\partial \m{x}_{d_{j}}}{\partial \m{x}_{d_{j-1}}}\\  \tfrac{\partial \m{x}_{d_{j}}}{\partial \m{x}_{a_{j-1}}}\end{bmatrix}&= \begin{bmatrix}
			\eye_{n_d} + \tilde{h} (\mF_{\m{x}_{d,j}} \frac{\partial \m{x}_{d_{j}}}{\partial \m{x}_{d_{j-1}}} + \mF_{\m{x}_{d,j-1}})\\ 
			\tilde{h}   (\mF_{\m{x}_{a,j}}\frac{\partial \m{x}_{a_{j}}}{\partial \m{x}_{a_{j-1}}}+\mF_{\m{x}_{a,j-1}})\\ \end{bmatrix},\\ 
\begin{bmatrix}
	\tfrac{\partial \m{x}_{a_{j}}}{\partial \m{x}_{d_{j-1}}} \\
	\tfrac{\partial \m{x}_{a_{j}}}{\partial \m{x}_{a{j-1}}}
\end{bmatrix} &= \begin{bmatrix}
			\tilde{h}   (\mG_{\m{x}_{d,j}}\frac{\partial \m{x}_{d_{j}}}{\partial \m{x}_{d_{j-1}}}+\mG_{\m{x}_{d,j-1}})\\ 
			\mu \eye_{n_a} + \tilde{h}   (\mG_{\m{x}_{a,j}}\frac{\partial \m{x}_{a_{j}}}{\partial 	\m{x}_{a_{j-1}}}+\mG_{\m{x}_{a,j-1}})\\ \end{bmatrix},
	\end{align}
\end{subequations}
then, the explicit representation of $\frac{\partial \m{x}_{d_{j}}}{\partial \m{x}_{d_{j-1}}}$ and $\frac{\partial \m{x}_{a_{j}}}{\partial \m{x}_{a{j-1}}}$ can be written as

\begin{subequations}\label{eq:TRAPmu-NDAE_partial_Implicit2}
	\begin{align}
		\begin{bmatrix}\tfrac{\partial \m{x}_{d_{j}}}{\partial \m{x}_{d_{j-1}}}\\  \tfrac{\partial \m{x}_{d_{j}}}{\partial \m{x}_{a_{j-1}}}\end{bmatrix} =&\begin{bmatrix}
			\begin{bmatrix}
				\eye_{n_d} - \tilde{h}  \mF_{\m{x}_{d,j}}
			\end{bmatrix}^{-1}	\begin{bmatrix}
				\eye_{n_d} + \tilde{h}  \mF_{\m{x}_{d,j-1}}
			\end{bmatrix}\\ 
			\tilde{h}  (\mF_{\m{x}_{a,j}}\frac{\partial \m{x}_{a_{j}}}{\partial \m{x}_{a_{j-1}}}+\mF_{\m{x}_{a,j-1}})\\ \end{bmatrix},\\ 
		\begin{bmatrix}
			\tfrac{\partial \m{x}_{a_{j}}}{\partial \m{x}_{d_{j-1}}} \\
			\tfrac{\partial \m{x}_{a_{j}}}{\partial \m{x}_{a{j-1}}}
		\end{bmatrix} = &  \begin{bmatrix}
			\tilde{h}   (\mG_{\m{x}_{d,j}}\frac{\partial \m{x}_{d_{j}}}{\partial \m{x}_{d_{j-1}}}+\mG_{\m{x}_{d,j-1}})\mu^{-1}\\  
			\begin{bmatrix}
				\m{\mu}_{n_a} - \tilde{h}   \mG_{\m{x}_{a,j}}
			\end{bmatrix}^{-1}\begin{bmatrix}
				\m{\mu}_{n_a} +\tilde{h}   \mG_{\m{x}_{a,j-1}}
			\end{bmatrix}\\ \end{bmatrix},
	\end{align}
\end{subequations}
where $\m{\mu}_{n_a} := \mu \eye_{n_a}$. Now, let for TI method, $\mA^{-}_{n_{d,j}} = (\eye_{n_d} - \tilde{h}  \mF_{\m{x}_{d,j}})$, $\mA^{+}_{n_{d,j}} = (\eye_{n_d} + \tilde{h}  \mF_{\m{x}_{d,j}})$, $\mA^{-}_{n_{a,j}} = (\m{\mu}_{n_a} - \tilde{h}  \mG_{\m{x}_{a,j}})$ and $\mA^{+}_{n_{a,j}} = (\m{\mu}_{n_a} + \tilde{h}  \mG_{\m{x}_{a,j}})$, such that $\mA_{n_{a,j}} $ and $\mA_{n_{a,j}}$ are invertible. Then, $\frac{\partial \m{x}_{j+1}}{\partial \m{x}_{j}}$ can be explicitly formulated and written for TI discretization as

\begin{equation}\label{eq:TRAPJac_mu-NDAE_partial}
	\begin{bmatrix}
		\frac{\partial \m{x}_{j}}{\partial \m{x}_{j-1}} 
	\end{bmatrix} \hspace{-0.1cm}=\hspace{-0.1cm}
	\begin{bmatrix}
		\begin{bmatrix}
			\mA^{-}_{n_{d,j}}
		\end{bmatrix}^{-1} 
		\begin{bmatrix}
			\mA^{+}_{n_{d,j-1}}
		\end{bmatrix}\\  
		\tilde{h} (\mF_{\m{x}_{a,j}}\begin{bmatrix}
			\mA^{-}_{n_{a,j}}
		\end{bmatrix}^{-1}\begin{bmatrix}
			\mA^{+}_{n_{a,j-1}}\end{bmatrix} +\mF_{\m{x}_{a,j-1}} )\\ 
		\tilde{h}  (\mG_{\m{x}_{d,j}}	\begin{bmatrix}
			\mA^{-}_{n_{d,j}}
		\end{bmatrix}^{-1} 
		\begin{bmatrix}
			\mA^{+}_{n_{d,j-1}}
		\end{bmatrix}+\mG_{\m{x}_{d,j-1}})\mu^{-1}\\ 
		\begin{bmatrix}
			\mA^{-}_{n_{a,j}}
		\end{bmatrix}^{-1}\begin{bmatrix}
			\mA^{+}_{n_{a,j-1}}
		\end{bmatrix}\\
	\end{bmatrix}.
\end{equation}

The partial derivative is now explicitly defined; as such we can now concatenate the Jacobian $\m J(\cdot)$ over observation horizon $\mr{N}_{o}$. 

\section{Partial Derivatives for Gear's discretization}\label{apdx:Jac_Gear}
To capture the system dynamics for Gear's discretization of order $k_g$, we realize that the formulation for the partial differential $\frac{\partial \m{x}_{j}}{\partial \m{x}_{j-1}}$ is not fully representative for BDF of order $k_g$ except when $k_g=1$, for reasons that will be obvious shortly. Building the Jacobian by differentiating the states with respect to $j - k_g$ instead of $j - 1$ allows to fully capture the system dynamics. Herein, we show how this formulation captures the full system dynamics by taking Gear's discretization order $k_g =  3$ as an example. Differentiating the system dynamics with respect to $\m{x}_{j-1}$ for $k_g =3$ yields the following 

\begin{equation}\label{eq:mu-NDAE-differentiation}
	\begin{split}
		\mE_{\mu}\frac{\partial \m{x}_{j}}{\partial \m{x}_{j-1}} =& \mE_{\mu}
		(\alpha_{(1)}\underbrace{\frac{\partial \m{x}_{j-1}}{\partial \m{x}_{j-1}}}_{\eye_{n}}+
		\alpha_{(2)}\underbrace{\frac{\partial \m{x}_{j-2}}{\partial \m{x}_{j-1}}}_{0}+
		\alpha_{(3)}\underbrace{\frac{\partial \m{x}_{j-3}}{\partial \m{x}_{j-1}}}_{0}) \\ 
		&+ \tilde{h} \eye_{n} \begin{bmatrix}
			\frac{\partial \m{f}(\m{x}_{j})}{\partial \m{x}_{j}}|_{\m{x}_j}\frac{\partial \m{x}_{j}}{\partial \m{x}_{j-1}}\\
			\frac{\partial \m{g}(\m{x}_{j})}{\partial \m{x}_{j}}|_{\m{x}_j}\frac{\partial \m{x}_{j}}{\partial \m{x}_{j-1}}\\
		\end{bmatrix}.
	\end{split}
\end{equation}

\begin{figure*}[t]
	\vspace{-0.28cm}
	\subfloat{\includegraphics[keepaspectratio=true,scale=1]{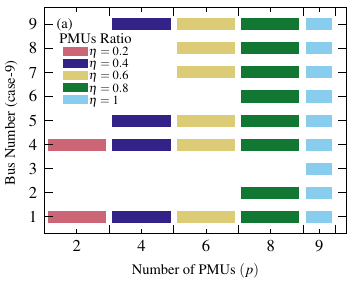}}
	\subfloat{\includegraphics[keepaspectratio=true,scale=1]{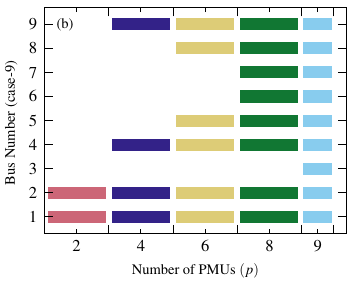}}
	\subfloat{\includegraphics[keepaspectratio=true,scale=1]{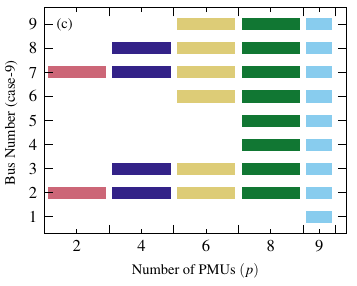}}
	\vspace{-0.5cm}
	\subfloat{\includegraphics[keepaspectratio=true,scale=1]{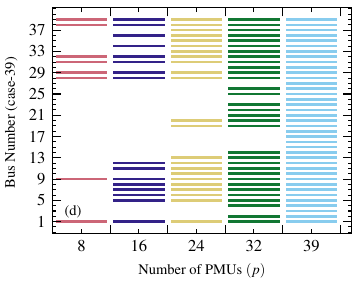}}
	\hspace{-0.1cm}\subfloat{\includegraphics[keepaspectratio=true,scale=1]{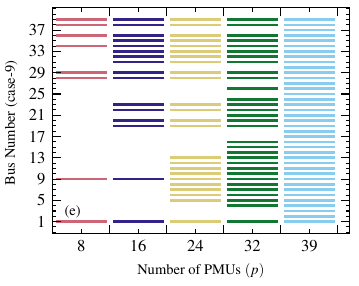}}
	\hspace{-0.1cm}\subfloat{\includegraphics[keepaspectratio=true,scale=1]{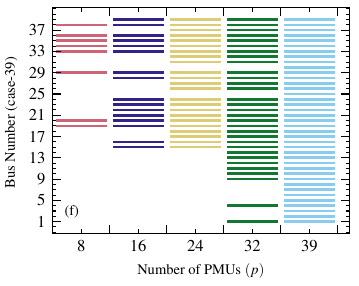}}
	\vspace{-0.5cm}
	\subfloat{\includegraphics[keepaspectratio=true,scale=0.97]{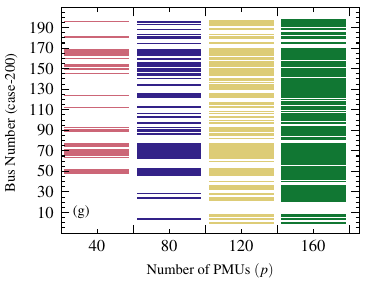}}
	\hspace{-0.1cm}\subfloat{\includegraphics[keepaspectratio=true,scale=0.97]{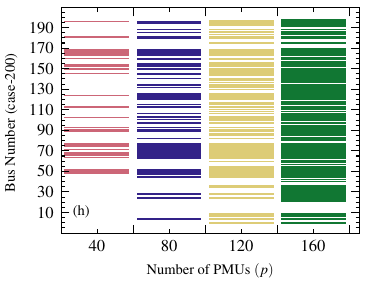}}
	\hspace{-0.1cm}\subfloat{\includegraphics[keepaspectratio=true,scale=0.97]{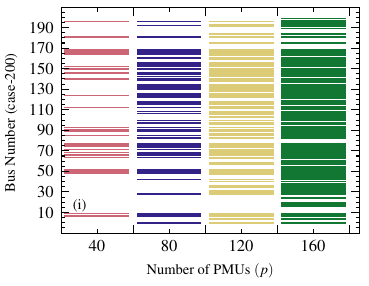}}
	\vspace{-0.2cm}
	\caption{Optimal PMU placement for $(a,b,c)$ $\mr{case}$-$\mr{9}$, $(d,e,f)$ $\mr{case}$-$\mr{39}$, and $(g,h,i)$ $\mr{case}$-$\mr{200}$ under different values of PMUs that are to be selected $p$: $(a,d,g)$ BE discretization, $(b,e,h)$ BDF discretization, and $(c,f,i)$ TI discretization. The figures depict, using horizontal bars, which PMUs are being selected for each of the PMU ratios $\eta$ being chosen. The width of the bars represent the total number of PMUs being added as compared to the previous ratio. The total number of PMUs being selected is represented by the PMU number $p$ on the x-axis. The location of the PMUs being selected is represented on the y-axis be referring to the bus number.}\label{fig:case_200-OPP}
	\vspace{-0.25cm}
\end{figure*}

From~\eqref{eq:mu-NDAE-differentiation} we notice that if we differentiate the system dynamics with respect to one prior time-step $(j-1)$ for $k_g>1$, the system dynamics depicted in the $\sum_{s=2}^{k_g}\alpha_s \frac{\partial \m{x}_{j-s}}{\partial \m{x}_{j-1}}$ become equal to zero. Utilizing such an approach, the BDF method for any order between $1\leq k_g\leq5$ will represent only that of order one. 

Based on such shortcoming, we consider the BDF discretization order for computing the partial derivatives. As such, we differentiate system~\eqref{eq:discrete_NDEA_concat} with respect to $k_g$ in order to capture the full system dynamics. For $k_g=3$, the partial derivative can be written as

\begin{equation}\label{eq:mu-NDAE-differentiation_kg}
	\begin{split}
		\mE_{\mu}\frac{\partial \m{x}_{j}}{\partial \m{x}_{j-3}} =& \mE_{\mu}
		(\alpha_{(1)}\frac{\partial \m{x}_{j-1}}{\partial \m{x}_{j-3}}+
		\alpha_{(2)}\frac{\partial \m{x}_{j-2}}{\partial \m{x}_{j-3}}+
		\alpha_{(3)}\underbrace{\frac{\partial \m{x}_{j-3}}{\partial \m{x}_{j-3}}}_{\eye_{n}})\vspace{-0.5cm} \\  
		&+ \tilde{h} \eye_{n} \begin{bmatrix}
			\frac{\partial \m{f}(\m{x}_{j})}{\partial \m{x}_{j}}|_{\m{x}_j}\frac{\partial \m{x}_{j}}{\partial \m{x}_{j-3}}\\
			\frac{\partial \m{g}(\m{x}_{j})}{\partial \m{x}_{j}}|_{\m{x}_j}\frac{\partial \m{x}_{j}}{\partial \m{x}_{j-3}}\\
		\end{bmatrix},
	\end{split}
\end{equation}
where the partial derivative $\frac{\partial \m{x}_{j-1}}{\partial \m{x}_{j-3}}= \frac{\partial \m{x}_{j-1}}{\partial \m{x}_{j-2}} \frac{\partial \m{x}_{j-2}}{\partial \m{x}_{j-3}}$.

For $j=k_g=3$, we now have defined $\frac{\partial \m{x}_{3}}{\partial \m{x}_{0}}$ as~\eqref{eq:mu-NDAE-differentiation_kg} and $\frac{\partial \m{x}_{2}}{\partial \m{x}_{0}} =\frac{\partial \m{x}_{2}}{\partial \m{x}_{1}} \frac{\partial \m{x}_{1}}{\partial \m{x}_{0}}$. Such that $\frac{\partial \m{x}_{2}}{\partial \m{x}_{1}}$ and $\frac{\partial \m{x}_{1}}{\partial \m{x}_{0}}$ are calculated using the formulation presented in~\eqref{eq:mu-NDAE_partial_Implicit3}.

Now, for $j \geq 2\times k_g$ that is in this case $j \geq 6$, we use the chain rule to compute the $j$-th derivative with respect to $\m{x}_0$ as follows
\begin{equation}\label{eq:Gear_Chainrule2}
	\frac{\partial \m{x}_{j}}{\partial \m{x}_0} = \frac{\partial \m{x}_j}{\partial \m{x}_{j-{k}_{g}}} 
	\cdots \frac{\partial \m{x}_{{k}_{g}}}{\partial \m{x}_0},
\end{equation}
and for $k_g<j<2 \times k_g$ that is in this case $3<j<6$, we represent the $j$-th derivative for $j=4$ and $j=5$ with respect to $\m{x}_0$ as follows
\begin{equation}\label{eq:Gear_Chainrule_4}
		\frac{\partial \m{x}_{4}}{\partial \m{x}_0} = \frac{\partial \m{x}_4}{\partial \m{x}_{1}} \frac{\partial \m{x}_{1}}{\partial \m{x}_0}, \;\;\;  \text{and}\;\;\;
		\frac{\partial \m{x}_{5}}{\partial \m{x}_0} = \frac{\partial \m{x}_5}{\partial \m{x}_{2}} 
		\frac{\partial \m{x}_{2}}{\partial \m{x}_1}\frac{\partial \m{x}_{1}}{\partial \m{x}_0}.
\end{equation}

Such that, $\frac{\partial \m{x}_{5}}{\partial \m{x}_{2}}$ and $\frac{\partial \m{x}_{4}}{\partial \m{x}_{1}}$ are calculated from~\eqref{eq:mu-NDAE-differentiation_kg}. Given such formulation, the Jacobian for Gear's discretization method is able to depict system dynamics for $k_g>1$.

\balance 
{\section{Proof of Proposition~\ref{prop:modularity}}\label{apdx:proof}
The following is the proof of Proposition~\ref{prop:modularity}.}
		\begin{proof}
			First we consider $\m W_o(\cdot)$ under full PMU placement, that is $\tilde{\m C} = \m C$. Then, for any $\mathcal{Z}\subseteq\mathcal{N}_{p}$, observe that the observability matrix can be written as
			\begin{subequations}\label{eq:proof-obs}
				\begin{align}
					\hspace{-0.3cm}	\m W_o(\cdot)=& \m{J}^T(\m{\Gamma},{\m x}_0)\m{J}(\m{\Gamma},{\m x}_0)\\
					=&\begin{bmatrix}
						\eye_{n} \\ \frac{\partial \m{x}_1}{\partial \m{x}_0} \\ 
						\vdots \\
						\frac{\partial \m{x}_{\mr{N}_{o}-1}}{\partial \m{x}_0}\\ 
					\end{bmatrix}^{\top }\hspace{-0.1cm}
					\begin{bmatrix}
						\eye_{\mr{N}_o} \kron \m{C}\end{bmatrix}^{\top}
					\begin{bmatrix}
						\eye_{\mr{N}_o} \kron \m{C}\end{bmatrix}
					\begin{bmatrix}
						\eye_{n} \\ \frac{\partial \m{x}_1}{\partial \m{x}_0} \\ 
						\vdots \\
						\frac{\partial \m{x}_{\mr{N}_{o}-1}}{\partial \m{x}_0}\\ 
					\end{bmatrix}.		
				\end{align}
			\end{subequations}
			
			We now reformulate~\eqref{eq:proof-obs} to show that it is analogous to a linear modular function. 	To do that, we refer to the distributive property of transpose over the Kronecker product. Thus, we can write $\begin{bmatrix} \eye_n \kron \m{C}\end{bmatrix}^{\top} = \begin{bmatrix} \eye_{n}^{\top} \kron \m{C}^{\top}\end{bmatrix}$,  we apply the dot-product rule to~\eqref{eq:proof-obs} as follows
			\begin{equation}\label{eq:proof-obs-2} 
				\m W_o(\cdot) =
				\sum_{k=0}^{\mr{N}_o-1}
				\bmat{ \frac{\partial \m{x}_{k}}{\partial \m{x}_0}}^{\top}
				\bmat{\m{C}^{\top} \m \Gamma^{2}\m{C}}
				\bmat{\frac{\partial \m{x}_{k}}{\partial \m{x}_0}},
			\end{equation}
			notice that $\m \Gamma^2= \m \Gamma$, since it is a binary matrix. Now, denoting $\m c_i\in\mbb{R}^{1\times n}$ as the $i$-th row of $\m C$ then
			\begin{subequations}
				\begin{align}
					\m W_o(\cdot) 
					&=
					\sum_{k=0}^{\mr{N}_o-1}
					\bmat{ \frac{\partial \m{x}_{k}}{\partial \m{x}_0}}^{\top}
					\left(\sum_{i=1}^{N_p} \gamma_i\m c_i^\top \m c_i \right)
					\bmat{\frac{\partial \m{x}_{k}}{\partial \m{x}_0}},\\
					&=
					\sum_{i=1}^{N_p} \gamma_i \left(\sum_{k=0}^{\mr{N}_{o}-1} 
					\bmat{ \frac{\partial \m{x}_{k}}{\partial \m{x}_0}}^{\top} 
					\bmat{\m c_i^\top \m c_i} 
					\bmat{\frac{\partial \m{x}_{k}}{\partial \m{x}_0}}\right),\\
					&=
					\sum_{i = 1}^{N_{p}} \m{W}_{o,i}(\mathcal{{\m{Z}}}_{i},\m{x}_0).
				\end{align}
			\end{subequations}
			
			Notice that for any $i\in \mathcal{Z}$ corresponds to a selected PMU, such that $\gamma_i = 1$. 	This shows that $\m{W}_{o}(\mathcal{Z},\m{x}_{0})$ is a linear matrix function and concurs with the definition of modularity\footnotemark[3]. Thus, the proof is complete.
\end{proof}

{\section{Results of the proposed OPP.}}\label{apdx:OPP}
The optimal PMU placements resulting from solving \textbf{P5} for the three test systems, under different PMU selection ratios and discretization methods, are depicted in Fig.~\ref{fig:case_200-OPP}.
\end{document}

%% file: preamble.tex
\usepackage{graphicx}
\usepackage{textcomp}
\usepackage{epsfig} 
\usepackage{epsfig,color,amsmath,cite}

\usepackage{amsthm} 
\usepackage{amsmath}    

\usepackage{enumitem} 
\setlist[itemize]{leftmargin=*}

\usepackage{bm}
\usepackage{epstopdf}
\usepackage{amssymb}
\usepackage{url}

\usepackage{multirow}
\usepackage{hhline}
\usepackage{booktabs}
\usepackage[linesnumbered,boxed,commentsnumbered,ruled,vlined,longend]{algorithm2e}
\usepackage{comment}
\newcommand{\eps}{\varepsilon}
\newcommand{\kron}{\otimes}

\DeclareMathOperator{\rank}{rank}

\DeclareMathOperator*{\minimize}{minimize}

\makeatother
\DeclareMathAlphabet\mathbfcal{OMS}{cmsy}{b}{n}

\newtheorem{mydef}{Definition}

\newtheorem{myrem}{Remark}

\newtheorem{myprs}{Proposition}



\usepackage{stackengine}

\newcommand{\mat}[1]{\boldsymbol{#1}}

\newcommand{\bmat}[1]{\begin{bmatrix} #1 \end{bmatrix}}

\providecommand{\eye}{\mat{I}}
\providecommand{\mA}{\ensuremath{\mat{A}}}

\providecommand{\mC}{\ensuremath{\mat{C}}}

\providecommand{\mE}{\ensuremath{\mat{E}}}
\providecommand{\mF}{\ensuremath{\mat{F}}}
\providecommand{\mG}{\ensuremath{\mat{G}}}

\providecommand{\mW}{\ensuremath{\mat{W}}}

\newcommand{\m}{\boldsymbol}
\allowdisplaybreaks[4]
\usepackage[colorlinks = true,
linkcolor = blue,
urlcolor  = blue,
citecolor = blue,
anchorcolor = blue]{hyperref}

\newcommand{\mbb}[1]{\mathbb{#1}}
\newcommand{\mr}[1]{\mathrm{#1}}
\usepackage[framemethod=TikZ]{mdframed}
\mdfdefinestyle{MyFrame}{%
	linecolor=black,
	outerlinewidth=1.25pt,
	roundcorner=1.25pt,
	innerrightmargin=5pt,
	innerleftmargin=5pt,}
	
\usepackage[noabbrev]{cleveref}

\usepackage{mathtools}

\DeclarePairedDelimiter\abs{\lvert}{\rvert}%
\DeclarePairedDelimiter\norm{\lVert}{\rVert}%

\makeatletter
\let\oldabs\abs
\def\abs{\@ifstar{\oldabs}{\oldabs*}}
\let\oldnorm\norm
\def\norm{\@ifstar{\oldnorm}{\oldnorm*}}
\makeatother

\usepackage[english]{babel}
\usepackage[utf8]{inputenc}
\usepackage[super]{nth}

\usepackage{graphicx}
\usepackage{float}
\usepackage[caption = false]{subfig}

\usepackage{array}
\usepackage{threeparttable}

\usepackage[english]{babel}
\usepackage[utf8]{inputenc}
\usepackage[super]{nth}

\usepackage{pifont}

\usepackage{mathtools}

\usepackage{mleftright}
\usepackage{xparse}

\NewDocumentCommand{\evaluat}{sO{\big}mm}{%
	\IfBooleanTF{#1}
	{\mleft. #3 \mright|_{#4}}
	{#3#2|_{#4}}%
}
